\documentclass[a4paper,fleqn,usenatbib]{mnras}

\usepackage{newtxtext,newtxmath}

\usepackage[T1]{fontenc}
\usepackage{ae,aecompl}

\usepackage{graphicx}	
\usepackage{amsmath}	
\usepackage{amssymb}	
\usepackage{lipsum}
\usepackage{pdflscape}  
\usepackage[dvipsnames]{color}
\usepackage[percent]{overpic}

\newcommand{\mrm}[1]{\mathrm{#1}}	
\newcommand{\vect}[1]{\boldsymbol{#1}}
\newcommand{\DRSun}[1]{$\mrm{DR_{\odot}}$} 
\newcommand{\ERSun}[1]{$\mrm{ER_{\odot}}$} 
\newcommand{\MFSun}[1]{$\mrm{MF_{\odot}}$} 
\newcommand{\ellsum}[1]{\ell_{\Sigma}} 

\title[Observing the simulations]{Observing the simulations: Applying ZDI to 3D non-potential magnetic field simulations}

\author[L. T. Lehmann et al.]{
L. T. Lehmann,$^{1}$\thanks{E-mail: ltl@st-andrews.ac.uk}
G. A. J. Hussain,$^{2,3}$
M. M. Jardine,$^{1}$
D. H. Mackay,$^{4}$
A. A. Vidotto$^{5}$
\\
$^{1}$SUPA, School of Physics and Astronomy, University of St Andrews, St Andrews KY16 9SS, UK \\
$^{2}$European Southern Observatory, Karl-Schwarzschild-Str. 2, 85748 Garching bei M\"unchen, Germany\\
$^{3}$Institut de Recherche en Astrophysique et Plan\'etologie, Universit\'e de Toulouse, UPS-OMP, 31400 Toulouse, France\\
$^{4}$School of Mathematics and Statistics, University of St Andrews, St Andrews KY16 9SS, UK\\
$^{5}$School of Physics, Trinity College Dublin, The University of Dublin, Dublin-2, Ireland\\
}

\date{Accepted XXX. Received YYY; in original form ZZZ}

\pubyear{2018}

\begin{document}
\label{firstpage}
\pagerange{\pageref{firstpage}--\pageref{lastpage}}
\maketitle

\begin{abstract}
The large-scale magnetic fields of stars can be obtained with the Zeeman-Doppler-Imaging (ZDI) technique, but their interpretation is still challenging as the contribution of the small-scale field or the reliability of the reconstructed field properties is still not fully understood.
To quantify this, we use 3D non-potential magnetic field simulations for slowly rotating solar-like stars as inputs to test the capabilities of ZDI. These simulations are based on a flux transport model connected to a non-potential coronal evolution model using the observed solar flux emergence pattern.
We first compare four field prescriptions regarding their reconstruction capabilities and investigate the influence of the spatial resolution of the input maps on the corresponding circularly polarised profiles. 
We then generate circularly polarised spectra based on our high resolution simulations of three stellar models with different activity levels, and reconstruct their large-scale magnetic fields using a non-potential ZDI code assuming two different stellar inclination angles. 
Our results show that the ZDI technique reconstructs the main features of slowly rotating solar-like stars but with $\sim$\,one order of magnitude less magnetic energy. The large-scale field morphologies are recovered up to harmonic modes $\ell \sim 5$, especially after averaging over several maps for each stellar model. While ZDI is not able to reproduce the input magnetic energy distributions across individual harmonic modes, the fractional energies across the modes are generally within 20\,\% agreement. The fraction of axisymmetric and toroidal field tends to be overestimated for stars with solar flux emergence patterns for more pole-on inclination angles.
\end{abstract}




\section{Introduction}
\label{Sec:Introduction}

The first detection of a magnetic field on a star other than the Sun was achieved by \cite{Babcock1947} for the Ap star 78\,Vir and for a solar-like star nearly 30 years later by \cite{Robinson1980}. Most detections of stellar magnetic fields are based on the Zeeman effect \citep{Zeeman1897}. The Zeeman broadening technique, \citep{Robinson1980,Reiners2006,Lehmann2015,Scalia2017}, compares the intensity profiles of magnetically sensitive with insensitive (i.e. effective Land\'e factor $g_{\rm eff} = 0$) spectral lines and determines the total average unsigned magnetic field of the observed star. Zeeman Broadening is sensitive to the mean field strength over the projected stellar disc (including both the large- and small-scale field) but provides no information about the distribution of the underlying magnetic morphology. Additionally, the strength of the broadening is small in relatively inactive cool stars like the Sun. As the Zeeman effect relies on measuring an additional source of broadening over rotational broadening, it becomes more challenging to measure in stars with high projected equatorial velocity $v_e \sin i$, and requires progressively higher levels of S/N \citep{Reiners2006}. 

The advent of large format spectropolarimeters covering the entire optical wavelength regime enabled circular polarisation signatures to be detected in cool stars systematically. The technique of Zeeman Doppler Imaging (ZDI)  \citep{Semel1989,Donati1997} was developed to reconstruct the large scale magnetic field morphologies in cool stars. It exploits the sensitivity of circular polarisation profiles to the longitudinal component of the stellar magnetic fields. ZDI relies on using the rotational modulation of high resolution circularly polarised spectra to reconstruct surface maps. The spatial resolution of these maps is limited by the rotation rate and inclination angle of the target star. \cite{Rosen2015} showed that the addition of linearly polarised spectra can provide additional constraints on the reconstructed maps. However, circularly polarised signatures are typically under 0.1\% of the continuum level, and linearly polarised spectra are typically an order of magnitude weaker than the circularly polarised signatures in cool stars (see e.g. \citealt{Donati1997,Rosen2015}). Therefore, linear polarisation is only accessible for the brightest most active stars, within a limited range of rotation periods. 

Over the last few decades the magnetic morphologies of tens of cool stars have been uncovered with ZDI. Several surveys including MagIcS\footnote{http://www.ast.obs-mip.fr/users/donati/magics/v1/}, Bcool\footnote{http://bcool.ast.obs-mip.fr/Bcool/Bcool\_\_\_cool\_magnetic\_stars.html}, MaTYSSE\footnote{https://matysse.irap.omp.eu/doku.php} and Toupies\footnote{http://ipag.osug.fr/Anr\_Toupies/} have observed stars of different ages, masses and Rossby numbers. The Rossby number $\mrm{Ro}$ is the ratio between the rotation period and the turnover time in the convection cell (see e.g. \citealt{Noyes1984}). The magnetic morphologies of cool stars can be split in three major groups: the slow rotators with Rossby numbers $\mrm{Ro} \ge 1$ , the faster rotators with lower Rossby numbers but stellar masses above $0.5\,\mrm{M_{\odot}}$ and the M dwarfs with masses below $0.5\,\mrm{M_{\odot}}$. The slowly rotating solar-like stars show weak, poloidal dominated and often very simple or dipolar morphologies \citep{Petit2008}. The fast rotators show stronger and often toroidal dominated fields that could appear as ring-like structures in the azimuthal field or as complex field morphologies \citep{Donati2008}. Mid-M~dwarfs near the fully convective boundary show very strong, poloidal but simple field morphologies, \citep{Morin2008a}, while late M~dwarfs display a bistable behaviour with both simple strong poloidal fields and weaker, toroidal, more complex morphologies \citep{Morin2010}. We recommend the review of \cite{Donati2009} for a more detailed overview of the trends found in the large-scale magnetic field morphologies of cool stars.

Many indirect tracers of magnetic activity (e.g. X-ray emission) increase with stellar rotation, \citep{Skumanich1972,Hartmann1987,Guedel2007,Reiners2012,Vidotto2014}. From ZDI maps of solar-type stars, \cite{Petit2008} found that  the toroidal fraction tends to increase with stellar rotation, and \cite{See2015} reported that the toroidal field scales more steeply with the inverse Rossby number than the poloidal field.

The first detection of a magnetic activity cycle was found from ZDI studies of $\tau$\,Boo, \citep{Donati2008a, Fares2009, Fares2013, Mengel2016, Jeffers2018}. The large-scale field is found to reverse its polarity every 120 days, which is much shorter than the 11-year cycle observed on the Sun \citep{Jeffers2018}. The continued monitoring of active stars from year to decadal timescales could uncover further magnetic cycles in the near future, adding to the long-term trends and cycles found for 61\,Cyg\,A \citep{BoroSaikia2015} and for $\epsilon$\,Eri \citep{Jeffers2014}. 

To detect magnetic field morphologies and cycles on solar-like stars -- stars with rotation periods longer than 10 days and activity levels similar to the Sun -- is especially challenging due to the slow rotation and relatively low activity levels. The magnetic activity of solar-like stars is weak compared to more active, rapidly rotating cool stars and long rotation periods restrict the spatial resolution of the large-scale magnetic field maps obtained with ZDI \citep{Petit2008}. 

Even after detecting the large-scale magnetic field morphology of a solar-like star, the comparison with the Sun is challenging. We observe the Sun with a very high spatial and temporal resolution and this prevents a direct comparison with cool star magnetic field maps reconstructed by ZDI. Through spherical harmonic decomposition, we can determine the large-scale field morphology of the Sun, enabling a direct comparison with the observed magnetic field morphologies of other stars \citep{Vidotto2016}.

The large-scale magnetic field morphology is still difficult to interpret as it misses most of the magnetic field in small-scale structures (see e.g. \cite{Lang2014}). Studies of M dwarfs by \cite{Reiners2009} and \cite{Morin2010} suggest that ZDI maps reconstruct between $6-14\,\%$ of the total magnetic field. The small-scale fields on stars other than the Sun are still unknown, but the scale sizes and distributions of small surface spots are starting to be uncovered thanks to exoplanet studies which use transiting planets as probes of the stellar surface structure \citep{Morris2017}. 

The question we want to answer in this paper is: how reliable are the detected magnetic field morphologies for slowly rotating solar-like stars and how should we interpret them? We use the 3D non-potential global magnetic field simulations of \cite{Gibb2016} as inputs to benchmark the ZDI technique. This is possible as the modulation of the solar and stellar magnetic fields, e.g. with flux transport models, is able to simulate most features of the observed solar magnetic field properties. \cite{Lehmann2018} showed that the large-scale field from the 3D non-potential simulations of \cite{Gibb2016} mimics the properties of observed solar-like stars, which makes them the perfect data set to test the reliability of ZDI and evaluate its ability to recover solar-like magnetic field morphologies. We determine what ZDI is able to recover and how to interpret the magnetic field maps for slowly rotating solar-like stars.

The paper is structured as follows: the simulations and techniques are explained in Section~\ref{Sec:SimulationsTechniques}. This includes a detailed overview of the ZDI technique and the assumptions and limitations inherent in the most common implementations of the technique in Section~\ref{Sec:ZDI}. We then describe our modelling and present the ZDI fits to these model Stokes profiles in Section~\ref{Sec:ModulationFittingStokesProfiles}. The results are displayed in Section~\ref{Sec:Results} followed by a discussion (Sec.~\ref{Sec:DiscussionConclusion}) and a summary in Section~\ref{Sec:Summary}.

\section{Simulations and Techniques}
\label{Sec:SimulationsTechniques}

\subsection{Simulated surface magnetic fields used in the ZDI tests}
\label{sec:simulations}
 
To test the reliability of the ZDI technique we use the 3D non-potential global magnetic field simulations of \cite{Gibb2016}. The highly resolved simulations are based on a magnetic flux transport model for the evolution of the photosphere, coupled with a non-potential evolution model for the corona out to $2.5\,\mrm{R_{\star}}$. To evolve the surface magnetic field the magnetohydrodynamic induction equation is solved, where a flux emergence pattern, usually bipolar starspot pairs, is advectively injected and sheared by the surface flux transport processes, i.e. surface differential rotation, poleward meridional flow and diffusion \citep{Charbonneau2014}. The field of the upper atmosphere responds to the surface processes by evolving through a series of non-linear force-free states by building up electric currents and free magnetic energy. 
Stellar flux emergence profiles and rates are unknown so we use the parametrised emergence model of \citet[sec.~3]{Gibb2016}. It is based on flux emergence properties determined by \cite{Yeates2014}, who used solar synoptic magnetograms from the US National Solar Observatory, Kitt Peak, during the activity maximum of cycle 23 from January 2000 to January 2001. \cite{Gibb2016} included the possibility to change certain parameters, e.g. flux emergence rate (ER) and differential rotation (DR). We use the simulations for stars with the following flux emergence $\mrm{ER} = 1,3,5\,\mrm{ER_{\odot}}$ and differential rotation rates $\mrm{DR} = 1,3,5\,\mrm{DR_{\odot}}$\footnote{We refer to sections 2 and 3 of \cite{Gibb2016} for the exact values of the solar flux emergence rate $\mrm{ER_{\odot}}$ and the solar differential rotation $\mrm{DR_{\odot}}$.}. For each star $\approx$\,$300$ vector magnetic maps are simulated over a time range of $\approx$\,$1$ year, with a spatial resolution of $0.9375^{\circ}$ at the equator. Full details of the simulations can be found in sections~2 and 3 of \cite{Gibb2016}.

Although the simulations provide the 3D vector magnetic field up to $2.5\,\mrm{R_{\star}}$, we only use the radial, azimuthal and meridional magnetic field map ($B_r, B_{\phi}, B_{\theta}$) of the photosphere in the present study. ZDI is sensitive to the magnetic field morphology of the stellar surface as the  technique has thus far only been applied to photospheric spectral lines. 
We select three stars from the simulations of \cite{Gibb2016}: the solar-like star ($\mrm{ER} = 1\,$\ERSun\ , $\mrm{DR} = 1\,$\DRSun\ ), a more active star ($\mrm{ER} = 3\,$\ERSun\ , $\mrm{DR} = 3\,$\DRSun\ ) and the most active star ($\mrm{ER} = 5\,$\ERSun\ , $\mrm{DR} = 5\,$\DRSun\ ). We choose ten maps per star, which are equally distributed across the simulated time range. Furthermore, we calculate the averaged properties of ten maps per star for the large- and small-scale magnetic field and compare them with the averaged results of all $300$ maps presented in \cite{Lehmann2018}. We ensure that the averages over the ten maps for each star have the same properties as the optimal average of the 300 maps per star.

\subsection{Extracting the large-scale field of the simulated maps}
\label{Subsec:ModellingTechniques}

The ZDI reconstructed surface maps are limited to a relatively low spatial resolution compared to the much higher resolution simulations. The difference in resolution is especially high for slowly rotating solar-like stars, as they have low $v_e \sin i$. The decomposition of the magnetic field morphology into its constituent spherical harmonics provides a fair order-of-magnitude comparison between the simulations and the observations, (see e.g. \citealt{Vidotto2016a,Lehmann2018}). 

By selecting a specific spherical harmonic mode $\ell$ of the decomposed magnetic field, one chooses a corresponding length scale, approximately described by $\theta \approx 180^{\circ}/\ell$. The large-scale field of the simulations can be filtered by selecting the low-order spherical harmonic modes, e.g. $\ell \leq 5$ or $\ell \leq 10$  (see e.g. \citealt{Morin2010}, \citealt{Johnstone2014}, \citealt{Yadav2015}, \citealt{Vidotto2016a}, \citealt{Folsom2016}, \citealt{Lehmann2017}).

The reconstructed magnetic field morphologies are often described in terms of their radial, azimuthal and meridional components, e.g., \cite{Morin2008a, Morin2010, Fares2012, Rosen2016}, or using the poloidal and toroidal component, e.g., \cite{Petit2008, Donati2009, See2015, Vidotto2016}. An additional property is the axisymmetric component, which governs the degree to which a particular magnetic field component is aligned with the rotation axis.

Following \cite{Elsasser1946} and \citet[Appendix III]{Chandrasekhar1961} the vector magnetic field $\vect{B}$ can be written using the associated Legendre polynomial $P_{\ell m} \equiv c_{\ell m}P_{\ell m}(\cos \theta)$ of mode $\ell$ and order $m$, where $c_{\ell m}$ is a normalization constant:
\begin{equation}
c_{\ell m} = \sqrt{\frac{2\ell+1}{4\pi}\frac{(\ell - m)!}{(\ell + m)!}}.
\end{equation}
The magnetic field can then be decomposed as follows for the radial, azimuthal and meridional component: 
\begin{align}
B_{r}(\theta, \phi) &= \sum_{\ell m} \alpha_{\ell m} P_{\ell m} e^{im\phi}, \label{Eq:B_rad} \\
B_{\phi}(\theta, \phi) &= - \sum_{\ell m} \beta_{\ell m} \frac{im P_{\ell m} e^{im\phi}}{(\ell + 1) \sin \theta} \nonumber \\ &+ \sum_{\ell m} \gamma_{\ell m} \frac{1}{\ell+1} \frac{\mathrm{d}P_{\ell m}}{\mathrm{d}\theta} e^{im\phi}, \label{Eq:B_azi}\\
B_{\theta}(\theta,\phi) &= \sum_{\ell m} \beta_{\ell m} \frac{1}{\ell+1} \frac{\mathrm{d}P_{\ell m}}{\mathrm{d}\theta} e^{im\phi} \nonumber \\ &+ \sum_{\ell m} \gamma_{\ell m} \frac{im P_{\ell m} e^{im\phi}}{(\ell + 1) \sin \theta}, \label{Eq:B_mer}
\end{align}
so that $(B_{r}, B_{\phi}, B_{\theta}) = \vect{B}$. The coefficients $\alpha_{\ell m}, \beta_{\ell m}$ and $\gamma_{\ell m}$ characterise the specific magnetic field morphology \citep[see][]{Donati2006a}. The radial field ($B_{r}$) points outwards, the azimuthal component ($B_{\phi}$) increases with longitude in the direction of rotation, and the meridional field ($B_{\theta}$) increases with colatitude from north to south. 
Alternatively, the magnetic field can also be decomposed into poloidal and toroidal field components: 
\begin{align}
B_{r,\mathrm{pol}}(\theta, \phi)  &= \sum_{\ell m} \alpha_{\ell m} P_{\ell m} e^{im\phi}, \nonumber \\
B_{\phi,\mathrm{pol}}(\theta, \phi)  &= - \sum_{\ell m} \beta_{\ell m} \frac{im P_{\ell m} e^{im\phi}}{(\ell + 1) \sin \theta}, \nonumber \\ 
B_{\theta,\mathrm{pol}}(\theta, \phi)  &=\sum_{\ell m} \beta_{\ell m} \frac{1}{\ell+1} \frac{\mathrm{d}P_{\ell m}}{\mathrm{d}\theta} e^{im\phi} , \label{Eq:B_pol}
\end{align}
\begin{align}
B_{r,\mathrm{tor}}(\theta, \phi) &= 0, \nonumber \\
B_{\phi,\mathrm{tor}}(\theta, \phi) &= \sum_{\ell m} \gamma_{\ell m} \frac{1}{\ell+1} \frac{\mathrm{d}P_{\ell m}}{\mathrm{d}\theta} e^{im\phi}, \nonumber \\
B_{\theta,\mathrm{tor}}(\theta, \phi) &= \sum_{\ell m} \gamma_{\ell m} \frac{im P_{\ell m} e^{im\phi}}{(\ell + 1) \sin \theta}, \label{Eq:B_tor}
\end{align}
so that $(B_{r,\rm{pol}}, B_{\phi,\rm{pol}}, B_{\theta,\rm{pol}}) = \vect{B_{\rm{pol}}}$, $(B_{r,\rm{tor}}, B_{\phi,\rm{tor}}, B_{\theta,\rm{tor}}) = \vect{B_{\rm{tor}}}$ and $(\vect{B_{\mrm{pol}}}, \vect{B_{\mrm{tor}}}) = \vect{B}$. The poloidal field is characterised by two coefficients $\alpha_{\ell m}$ and $\beta_{\ell m}$, while the toroidal field is only characterised by one coefficient $\gamma_{\ell m}$.
The sums run from $1\leq \ell \leq \ell_{\mathrm{max}}$ and from $\vert m \vert \leq \ell$, where $\ell_{\mathrm{max}}$ is the selected maximum mode of the spherical harmonic decomposition corresponding to the smallest included length scale. The axisymmetric modes are the modes where $m=0$. Using these equations, we are able to decompose every vector magnetic field into different length-scales, applying the algorithms published by \cite{Vidotto2016a}. 

For further analysis we can determine the mean-squared flux density per field component, e.g. for the radial field:
\begin{equation}
\langle B^2_{r}\rangle = \tfrac{1}{4\pi}\textstyle \int \textstyle B^2_{r}(\theta, \phi)\sin(\theta)\,\mathrm{d}\theta \mathrm{d}\phi.
\end{equation}
In the following we will call the mean squared flux density $\langle B^2\rangle\mathrm{[G^2]}$, the magnetic energy. The mean-squared flux density is a good proxy for the magnetic energy of the simulations even if it is not exactly equivalent. For the reconstructed maps, $\langle B^2\rangle\mathrm{[G^2]}$ is also often called the magnetic energy, even though it is restricted to the net magnetic flux per resolution element, e.g. see review of \cite{Reiners2012}.
The determination of the magnetic energy is either possible for a specific length-scale by choosing the corresponding $\ell$-mode or for a specific sub-structure, e.g. the large-scale field, by choosing the corresponding cumulative $\ellsum\ $-modes. The cumulative modes $\ellsum\ $ include all lower modes, e.g. the cumulative $\ellsum\ \leq 5$ includes all $\ell$-modes $\ell = 1$ to $\ell = 5$.

\subsection{Zeeman Doppler Imaging}
\label{Sec:ZDI}

Doppler imaging techniques have been used to reconstruct the brightness and abundance distributions across the surfaces of cool stars and chemically peculiar Ap stars for over 30 years \citep{VogtPenrodHatzes1987}. 
The technique relies on each area of the visible stellar disc producing a specific Doppler-shifted contribution to the corresponding photospheric line profile. Spotted regions that are unevenly distributed across the stellar disc cause Doppler-shifted distortions in photospheric line profiles that are modulated with the rotation period of the star. These modulations depend on the longitude and latitude of the region and how it is oriented on the projected stellar disc as the star rotates. The more rapidly rotating the star, the greater its $v_e \sin i$, and the greater the range of Doppler-shifted contributions across the stellar surface; this, in turn, increases the amount of spatial information encoded in the corresponding spectral line profiles. High rotation rates do however pose an additional challenge in that rotationally broadened absorption line profiles become increasingly shallow. Doppler imaging campaigns of rapid rotators therefore have the additional challenge of acquiring data with sufficiently high S/N to detect structure robustly in the broadened line profiles while maintaining small enough exposure times to ensure minimal phase-blurring. 

In practice, while it is relatively simple to transform from image to data space, the reverse transformation is an ill-posed problem with a large number of spot distributions able to fit a particular dataset within a specified level of fit. Variable data quality in the spectral time-series and gaps in phase coverage further exacerbate the problem. A regularising function is therefore required to obtain a unique solution. 

Different implementations of ZDI have experimented with different regularising functions, e.g., different implementations of maximum entropy \citep{Donati1997, Hussain2000}, Levenberg-Marquardt minimisation constrained by Tikhonov regularisation \citep{Piskunov2002}, and most recently \cite{Carroll2012} report an implementation of an iterative Landweber method which aims to minimise the sum of the squared errors. We focus on the use of maximum entropy in this study. With maximum entropy regularisation, the image with the least amount of information required to fit the data to a specified $\chi^2$ value is selected.

\cite{Semel1989} proposed the application of Doppler imaging principles to circularly polarised (Stokes V) profiles to recover the large-scale magnetic fields on the surfaces of magnetically active stars. As with Doppler imaging, this technique -- called Zeeman Doppler Imaging (ZDI) -- enables the locations of magnetically active regions to be inferred by tracing the rotationally modulating Stokes V signatures in a spectral time-series. ZDI can also recover the vector information of the large scale magnetic field as Stokes V profiles are sensitive to the longitudinal component of the magnetic field. The rotationally modulated Stokes V signatures in a full timeseries enable us not only to pinpoint the location but the orientation of the field. 

Stokes V signatures are very weak in cool stars -- even in relatively active stars, the polarisation is typically at $\sim$0.1\% of the continuum level. It is therefore usually necessary to use a large format spectropolarimeter that covers a significant fraction of the visible spectrum in order to collect polarisation information across thousands of photospheric lines. The optical spectro-polarimeters ESPADONS (at the CFHT), NARVAL (at TBL) and HARPS in polarimetric mode (on the ESO 3.6-metre telescope) are the most commonly used instruments to detect and map stellar magnetic fields using circularly polarised Stokes V spectra. Additionally, it is necessary to use a cross-correlation technique, summing up the signature contained in thousands of photospheric profiles across the entire large format spectrograph to recover a robust signature with enhanced S/N (at a level of several thousand). The two main multi-line techniques currently used to recover robust Stokes V signatures in cool stars are Least Squares Deconvolution \citep{Donati1997b} and Single Value Decomposition \citep{Carroll2012}.

The first implementations of ZDI tended to decompose the surface magnetic field into three independent field vectors: radial, azimuthal (east-west) and meridional (north-south) vectors \citep{Donati1997c,Hussain2000}. The first targets of ZDI were rapidly rotating stars, with $v_e \sin i$ values of between 50--90$\,\rm{km/s}$. \cite{Petit2002} demonstrated that the unique properties of Stokes V signatures also enabled large scale field maps to be reconstructed for stars with $v_e \sin i< 15\,\rm{km/s}$; these tend to be more slowly rotating stars for which brightness maps could not be reconstructed with conventional Doppler imaging, as their starspot activity is generally lower and has a correspondingly smaller impact on the narrow absorption line profiles. While the Stokes V signatures are very weak in these slowly rotating, Sun-like stars compared to their faster rotating and more magnetically active counterparts, the maps recovered can display a range of large scale field morphologies \citep{Petit2002}.

\subsubsection{Limitations of ZDI}
\label{sec:limitations}

In this section we summarise the main assumptions and limitations inherent when applying ZDI to stars with solar like activity levels, highlighting which of these are explicitly addressed by our study.
\begin{itemize}

\item As Stokes V profiles are only sensitive to the longitudinal component of the surface magnetic field, they are prone to flux cancellation, particularly for stars with solar-like activity levels, in which flux is expected to emerge in bipolar magnetic region pairs by analogy with the Sun. 
What exactly is being reconstructed in ZDI maps remains an open question, and the effects of flux cancellation are expected to worsen in more slowly rotating stars. 

\item The stellar inclination angle combined with the sensitivity of Stokes V profiles to the longitudinal component of the magnetic field can lead to cross-talk between the radial and meridional field components. \cite{Donati1997} show that this can affect the reconstructions of the magnetic field at low latitudes, particularly in stars with low inclination angles (i.e., that are more pole-on).

\item As noted above, the first implementations of ZDI mapped the large scale field by decomposing it into its constituent radial, azimuthal and meridional field components. However, these solutions could easily reconstruct physically unrealistic solutions (e.g., monopoles). Additionally, simulations revealed that simple configurations, such as dipole fields could not be recovered. Therefore, a spherical harmonics description has been adopted by most codes that are currently in use (\citealt{Hussain2002,Donati2006,Kochukhov2015}). Different spherical harmonic descriptions of the surface field are detailed in Section\,\ref{sec:shdescription}.

\item The inclination angle of the star has a clear impact on the reconstructed maps. For stars that are seen almost equator-on, due to the Doppler-shifted contributions from the northern and southern hemispheres being identical, there is often mirroring about the equator in the reconstructed maps. For pole-on stars, conversely, as there is no contribution from most of the (obscured) stellar surface, this raises questions of how to fill in the missing flux. By defining the surface magnetic field in terms of spherical harmonics, it is possible to start to fill in the field in the missing hemisphere (assuming $\operatorname{div} \vect{B} = 0$). In order to investigate whether an antisymmetric or symmetric field is more likely, weights can be added to the corresponding $\ell$ modes. This has been done very successfully to explain the polar accretion flows seen in accreting pre-main sequence stars \citep{Donati2011}. 

\item All G-type stellar surfaces are expected to possess a combination of cool dark spots and bright facular regions (which tend to be brightest on the limb). This uneven brightness distribution can affect the relative strengths of the circular polarisation profiles across these regions; indeed, the contributions to the circularly polarised profiles from the strongest umbral features are likely not being detectable at all in the photospheric line profiles. In rapid rotators, magnetic and brightness maps can be reconstructed simultaneously as inhomogeneous brightness distributions can be detected down to a  spatial resolution that is limited by the spectral resolution of the data, and the star's $v_e \sin i$. However, in slowly rotating solar-type stars that are the subject of this study, it is impossible to recover the brightness distributions as starspots rotating across the stellar surface have an almost imperceptible effect on the time-series of intensity line profiles. 

\item The impact of variable quality and S/N may also affect the integrity of ZDI reconstructions, particularly when comparing maps for the same star taken at various stages of its activity cycle (e.g., $\tau$ Boo; \citealt{Mengel2016}). To address this problem and to ensure consistent results across datasets of different quality for the same star, a robust stopping criterion, such as that proposed by \cite{Alvarado-Gomez2015} would be essential. The minimum useful phase coverage likely depends on the maximum attainable resolution of the reconstructed map, which as mentioned earlier also depends on the $v_e \sin i$ of the star.
\end{itemize}

\subsubsection{Comparing the different descriptions of the magnetic field}
\label{sec:shdescription}

%
%
%
\begin{figure*}
\raggedright
\large
\hspace{1.8cm}\textbf{Input map}\hspace{3.3cm}\textbf{Potential $\alpha = \beta$}\hspace{2.9cm}\textbf{Original map}\\
\hspace{1.4cm}(Visible surface) \hspace{8.1cm} (Complete surface)\\
\centering
\includegraphics[angle=0,width = \textwidth ,clip]{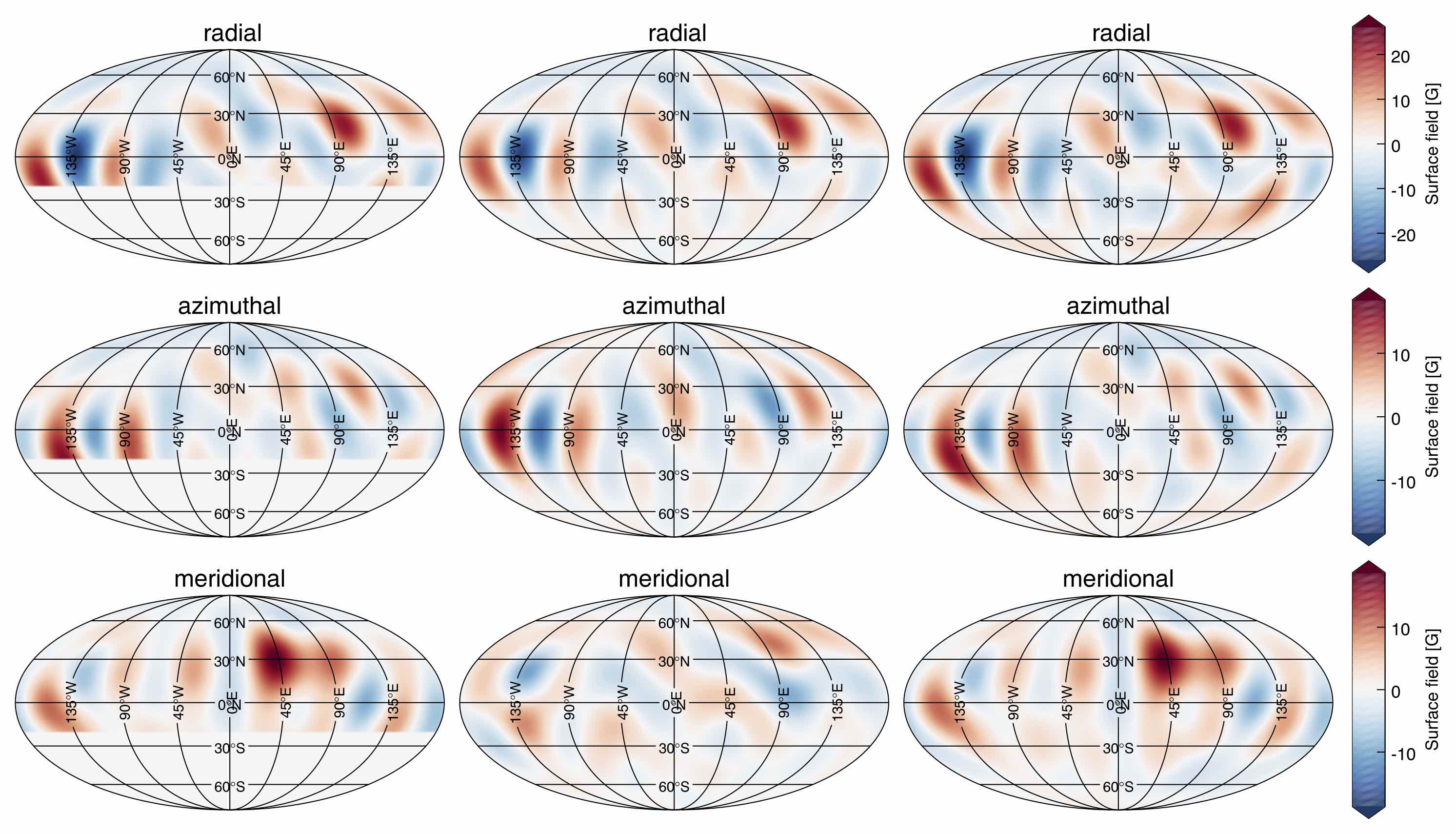}
\caption{The Mollweide projected magnetic field maps of the different descriptions of the magnetic field for the solar-like star ($\mrm{ER} = 1\,$\ERSun\ \ and $\mrm{DR} = 1\,$\DRSun\  ). The radial component is displayed in the \textit{top row}, the azimuthal in the \textit{middle row} and the meridional component at the \textit{bottom row}. From \textit{left to right}: the input map with a restricted large-scale field to $\ell_{\Sigma} = 7$  and truncated latitudes corresponding to an inclination of $i=20^{\circ}$, the potential ($\alpha_{\ell m} = \beta_{\ell m}$) reconstruction and the original simulated maps for $\ell_{\Sigma} = 7$.}
\label{Fig:DecompMap11_1930}
\end{figure*}

The Zeeman-Doppler-Imaging techniques can use different descriptions of the large-scale field to reconstruct the stellar magnetic field. They have been tested and compared for simple magnetic fields or single spot configurations (see e.g. \citealt{Donati1997, Hussain2001}) or by fitting the observed Stokes V profiles (see e.g. \citealt{Hussain2002, Donati2006, Kochukhov2016a}). 

In this section, we compare four different descriptions of the magnetic field using as input data the non-potential 3D magnetic field simulations. At this point we are not producing ZDI reconstructions using these different field descriptions. We are testing the reconstruction abilities of the different field descriptions directly on our simulations. This allows us to investigate which descriptions are most effective when reconstructing the magnetic field morphology of our different surface vector magnetic field simulation, without considering any other aspect of the ZDI.

We compare four different descriptions based on the spherical harmonic decomposition, including two potential and two non-potential models. We have not taken into account the direct ZDI approach that fits independent magnetic structures directly to the Stokes profiles. The direct ZDI approach causes non-physical $\operatorname{div} \vect{B} \neq 0$ field configurations, as the different field components are not related to each other. We test therefore only the descriptions that are based on the spherical harmonic decomposition, where the magnetic field components are related to each other and follow Maxwell's Equations. The spherical harmonic decomposition, see Eq.~\ref{Eq:B_rad}-\ref{Eq:B_mer} in Section~\ref{Subsec:ModellingTechniques}, uses the coefficients $\alpha_{\ell m}, \beta_{\ell m}$ and $\gamma_{\ell m}$ to specify the field structure. The different descriptions require different coefficients. The potential model requires $\gamma_{\ell m} = 0$, which leads to a purely poloidal field configurations as the toroidal field $\vect{B}_{\mrm{tor}} = 0$ (see Eq.~\ref{Eq:B_tor}). The non-potential models allows $\gamma_{\ell m} \neq 0$, so that $\gamma_{\ell m}$ can be fitted to the field structure and $\vect{B}_{\mrm{tor}} \neq 0$. Further descriptions force $\alpha_{\ell m} = \beta_{\ell m}$ as this allows the extrapolation of the magnetic field from the stellar surface to the corona, \citep{Hussain2002, Jardine2013} but this approach reduces the degree of freedom, as also seen with the potential models.

We compare the following descriptions of the magnetic field:
\begin{itemize}
\item \textbf{potential $\alpha_{\ell m} = \beta_{\ell m}$}: This model has the fewest degrees of freedom as $\alpha_{\ell m} = \beta_{\ell m}$ and $\gamma_{\ell m} = 0$. 
\item \textbf{potential $\alpha_{\ell m} \neq \beta_{\ell m}$}: This model gains one degree of freedom by allowing $\alpha_{\ell m} \neq \beta_{\ell m}$ but still requires $\gamma_{\ell m} = 0$. It should be able to better fit the meridional and azimuthal field components (see \citealt{Hussain2001}).
\item \textbf{non-potential $\alpha_{\ell m} = \beta_{\ell m}$}: This model allows $\gamma_{\ell m} \neq 0$ and therefore a toroidal field component by providing the possibility to extrapolate the field structure from the stellar surface to higher atmospheres (see \citealt{Hussain2002}).
\item \textbf{non-potential $\alpha_{\ell m} \neq \beta_{\ell m}$}: This model has the maximum number of degrees of freedom by allowing $\alpha_{\ell m} \neq \beta_{\ell m}$ and $\gamma_{\ell m} \neq 0$ and should perfectly reconstruct the magnetic field structure at the surface (see e.g. \citealt{Donati2006}). 
\end{itemize} 

We examine the simulated large-scale field of all 30 stellar maps for the cumulative $\ellsum\ \leq 7$ modes. The latitudes that are invisible due to the inclination effect are truncated, see Fig.~\ref{Fig:DecompMap11_1930}-\ref{Fig:DecompMap55_2113_60}, \textit{left column}. The input map represents therefore the maximum possible field that is observable with ZDI. We reconstruct the magnetic field map for $\ellsum\ \leq 7$ using the \cite{Vidotto2016} method based on the Eq. \ref{Eq:B_rad}-\ref{Eq:B_mer} by altering the coefficients $\alpha_{\ell m}, \beta_{\ell m}$ and $\gamma_{\ell m}$ depending on the model, see Fig.~\ref{Fig:DecompMap11_1930}-\ref{Fig:DecompMap55_2113_60}, \textit{middle columns}. We compare the reconstructed magnetic field of the different models with the original map (simulated large-scale field $\ellsum\ \leq 7$ without truncated latitudes), see Fig.~\ref{Fig:DecompMap11_1930}-\ref{Fig:DecompMap55_2113_60}, \textit{right column}. Additional figures using the same format can be found in the appendix, see Fig.~\ref{Fig:DecompMap11_2065}-\ref{Fig:DecompMap33_1992}. We further compute the correlation coefficient between the reconstructed maps and the original map.

In general, we find that the non-potential $\alpha_{\ell m} \neq \beta_{\ell m}$ model is always able to reconstruct the original map to a satisfactory, i.e. level showing the highest correlation coefficients. The spherical harmonics description even predicts the magnetic field morphology in the invisible (obscured) hemisphere to a certain extent, which gives a higher agreement between reconstructed  and original maps than the input  and original maps. The other models often miss essential field structures. The potential $\alpha_{\ell m} = \beta_{\ell m}$ is the worst as expected from the degrees of freedom. From a mathematical point of view, the first three descriptions are not able to reconstruct the input magnetic field. The limitations of $\alpha_{\ell m}, \beta_{\ell m}$ and $\gamma_{\ell m}$ prevent the reconstruction of essential structures of the input magnetic field morphologies, so that no ZDI code using these descriptions will be able to reconstruct the correct field morphology of these input maps. Nevertheless, we find that some of the simulated magnetic field morphologies are able to be reconstructed to a convincing level by the more restricted models.

Specifically, the radial component is the same for all four models as the reconstruction of the radial field only depends on the coefficient $\alpha_{\ell m}$, see Eq.~\ref{Eq:B_rad}. The radial field is therefore always equally well reconstructed regardless of the description and reaches correlation coefficients higher than 0.98 for inclination $60^{\circ}$ and 0.9 for inclination $20^{\circ}$. 
The azimuthal field component can be reconstructed with the potential models for some of the solar case simulations. Fig.~\ref{Fig:DecompMap11_1930} shows an example, where the azimuthal field (\textit{middle row}) reconstruction is acceptable for the potential $\alpha_{\ell m} = \beta_{\ell m}$ model. Fig.~\ref{Fig:DecompMap11_2065} in the appendix displays an example where the potential $\alpha_{\ell m} = \beta_{\ell m}$ model is no longer able to fit the azimuthal component, but the potential $\alpha_{\ell m} \neq \beta_{\ell m}$ model is. As the stars become more active (increasing the flux emergence rate and differential rotation) the potential model becomes increasingly inadequate. The correlation coefficients for both potential models decrease while the correlation coefficients for both non-potential models increase with the activity level of the star. Fig.~\ref{Fig:DecompMap33_1992} in the appendix shows an example for a more active star ($\mrm{ER} = 3\,$\ERSun\ \ and $\mrm{DR} = 3\,$\DRSun\  ), where the potential models are no longer acceptable, but the non-potential $\alpha_{\ell m} = \beta_{\ell m}$ model is. For the most active stars ($\mrm{ER} = 5\,$\ERSun\ \ and $\mrm{DR} = 5\,$\DRSun\  ) only the non-potential $\alpha_{\ell m} \neq \beta_{\ell m}$ model can fit the azimuthal input map, see Fig.~\ref{Fig:DecompMap55_2113_20}. 
To reconstruct the meridional component the non-potential $\alpha_{\ell m} \neq \beta_{\ell m}$ is often the only satisfactory model. The correlation coefficient decreases by 10-20\,\% per degree of freedom, independent of the activity level of the star. We find that the meridional component benefits from allowing $\alpha_{\ell m} \neq \beta_{\ell m}$, see Fig.~\ref{Fig:DecompMap55_2113_20} \textit{bottom row}. 

The inclination also affects the reconstruction as depending on the inclination angle a part of the southern hemisphere is hidden. The correlation coefficients increase by 10-20\,\% from inclination $20^{\circ}$ to $60^{\circ}$. The field reconstructions for the more active stars are often widely affected by the inclination effect. The large-scale fields of the active stars mainly consist of two azimuthal band-like structures at mid latitudes with opposite polarity on the two hemispheres, see Fig.~\ref{Fig:DecompMap55_2113_60}. At low inclination the structure of the southern band is missing, see Fig.~\ref{Fig:DecompMap55_2113_20}, so that the true large-scale morphology is hidden from the observer.

\begin{landscape}

%
%
%
\begin{figure}
\raggedright
\large
\hspace{1.9cm}\textbf{Input map}\hspace{1.2cm}\textbf{Potential $\alpha = \beta$}\hspace{0.8cm}\textbf{Potential $\alpha \neq \beta$}\hspace{0.4cm}\textbf{Non-potential $\alpha = \beta$}\hspace{0.1cm}\textbf{Non-potential $\alpha \neq \beta$}\hspace{0.3cm}\textbf{Original map}\\
\hspace{1.6cm}(Visible surface) \hspace{14.3cm} (Complete surface)\\
\centering
\includegraphics[angle=0,width = 1.2\textheight ,clip]{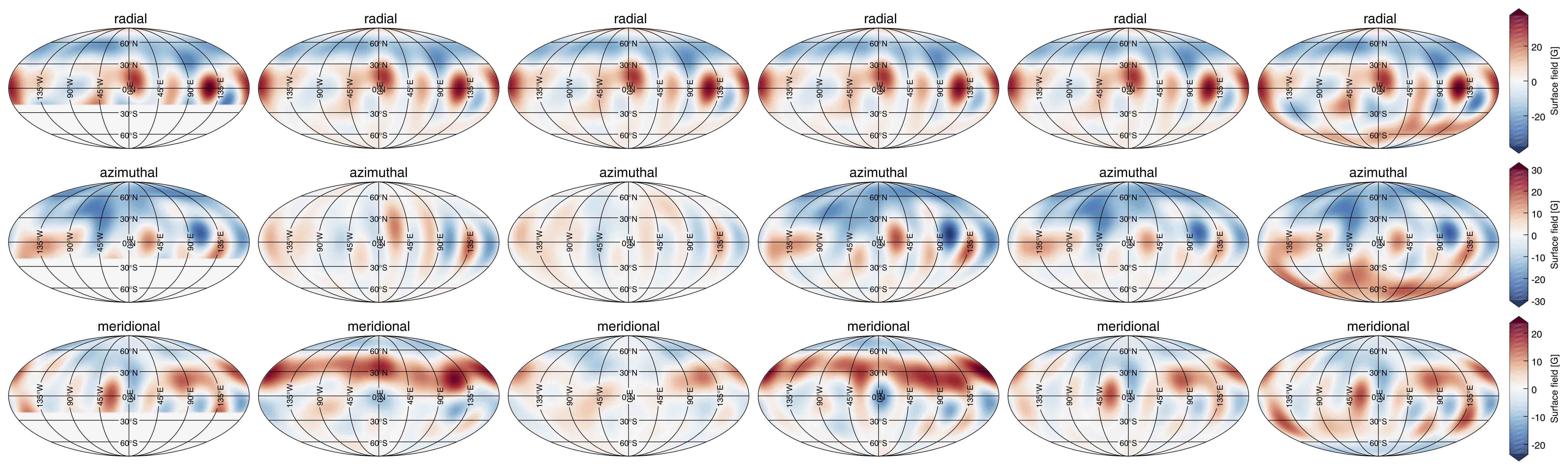}
\caption{The Mollweide projected magnetic field maps of the different descriptions of the magnetic field for the most active star ($\mrm{ER} = 5\,$\ERSun\ \ and $\mrm{DR} = 5\,$\DRSun\  , $i=20^{\circ}$). From \textit{left to right}: the input map including $\ell_{\Sigma} = 7$  and truncated latitudes corresponding to an inclination of $i=20^{\circ}$, the potential ($\alpha_{\ell m} = \beta_{\ell m}$), the potential ($\alpha_{\ell m} \neq \beta_{\ell m}$), the non-potential ($\alpha_{\ell m} = \beta_{\ell m}$), the non-potential ($\alpha_{\ell m} \neq \beta_{\ell m}$) reconstruction and the original simulated maps for $\ell_{\Sigma} = 7$. The format is the same as in Fig.~\ref{Fig:DecompMap11_1930}.}
\label{Fig:DecompMap55_2113_20}
\end{figure}

%
%
%
\begin{figure}
\raggedright
\large
\hspace{1.9cm}\textbf{Input map}\hspace{1.2cm}\textbf{Potential $\alpha = \beta$}\hspace{0.8cm}\textbf{Potential $\alpha \neq \beta$}\hspace{0.4cm}\textbf{Non-potential $\alpha = \beta$}\hspace{0.1cm}\textbf{Non-potential $\alpha \neq \beta$}\hspace{0.3cm}\textbf{Original map}\\
\hspace{1.6cm}(Visible surface) \hspace{14.3cm} (Complete surface)\\
\centering
\includegraphics[angle=0,width = 1.2\textheight ,clip]{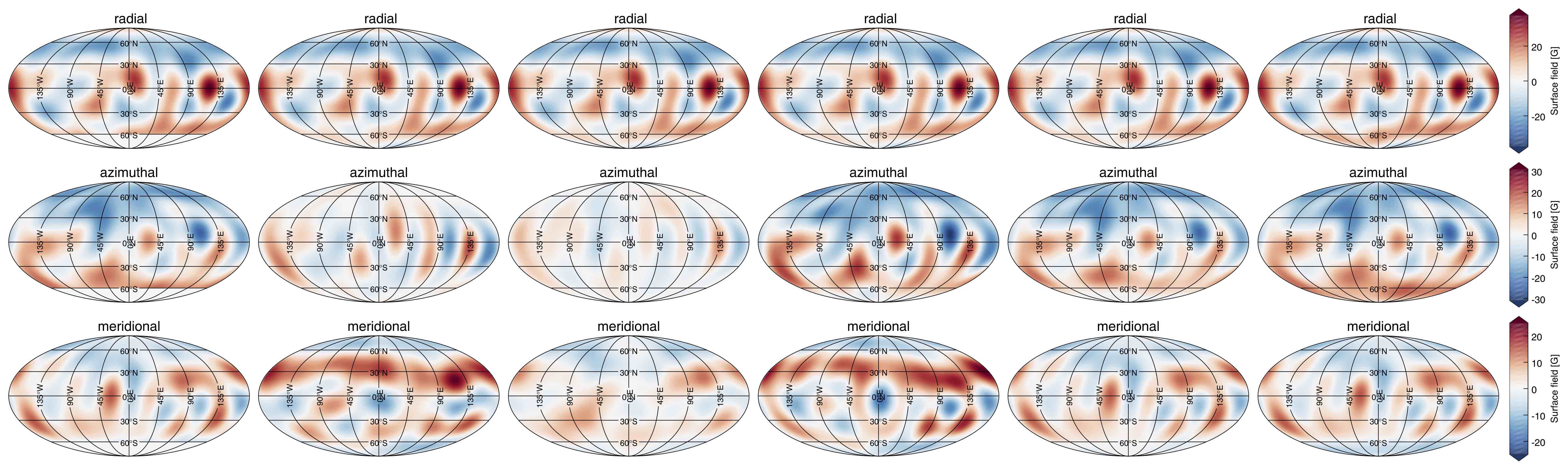}
\caption{Mollweide projected magnetic field maps of the different descriptions of the magnetic field for the same star and map as in Fig.~\ref{Fig:DecompMap55_2113_20} but for a higher inclination angle $i=60^{\circ}$ ($\mrm{ER} = 5\,$\ERSun\ \ and $\mrm{DR} = 5\,$\DRSun\  , $i=60^{\circ}$). The format is the same as in Fig.~\ref{Fig:DecompMap55_2113_20}.}
\label{Fig:DecompMap55_2113_60}
\end{figure}

\end{landscape}

\subsubsection{ZDI Testing: Our Implementation}

As described in Section~\ref{sec:limitations}, ZDI has a number of limitations. In particular, the spatial resolution of the reconstructed maps is determined by a combination of the $v_e \sin i$ of the star, the instrumental spectral resolution, along with the phase coverage, with the lowest spatial resolutions for stars with similar $v_e \sin i$ values as the Sun ($\sim 1.5\,\rm{km/s}$). We use the models described in Section~\ref{sec:simulations}
 to simulate realistic Stokes V datasets for these types of stars. We first compute a rotation period commensurate with the activity level in each of our selected models to derive the corresponding $v_e \sin i$ of our test datasets. 

For each of our simulated stars, we generate a time-series of Stokes V profiles using 25 phases that evenly sample the full stellar rotation period. Our reconstructions do not therefore investigate the effect of gaps in phase coverage and our reconstructions are carried out under the assumption that the large scale field does not change significantly over one stellar rotation period.

We investigate the effects of missing flux from the obscured hemisphere by considering two sets of stellar inclination angles: the first set of models has a 60$^{\circ}$ inclination angle -- i.e., there should be contributions to the Stokes V profiles from the large-scale field down to $\sim -30^{\circ}$ latitude. These are compared with reconstructions for stars that are significantly more pole on -- with inclination angles of  20$^{\circ}$ -- in which the signal is dominated by the polar regions. 

The aim of our tests is to determine how much of the large-scale field can be robustly reconstructed in solar-type stars, given the issues with flux cancellation and obscured hemisphere outlined earlier. We explore how the reconstructed field compares to the field in our input models, including a detailed comparison between the global properties in the reconstructed field and the properties of the large scale field in the input models. All spherical harmonic modes that are allowed are essentially weighted equally so the reconstructions test the effect that the maximum entropy regularisation has when applied to the spherical harmonic descriptions described in the previous section. 

As our models are limited to slowly rotating stars with solar-type activity levels, the  varying contribution from rotating spot and facular regions does not leave a detectable signature in the intensity profiles. Therefore our analysis is limited to a study of the magnetic field only, and cannot address the potential issue of inhomogeneous brightness distributions affecting the circularly polarised profiles described in Section~\ref{sec:limitations}.

\section{The Modulation and The Fitting of the Stokes Profiles}
\label{Sec:ModulationFittingStokesProfiles}

\subsection{Modelling the Stokes V profiles}

We model the Stokes~I and V profiles for each of the 10 randomly selected maps using three different stellar activity models at two inclination angles ($i=20^{\circ}$ and $60^{\circ}$). This gives a total of 60 time-series of Stokes~I and V spectral profiles. 

The local Stokes I profile is generated assuming an intrinsic line profile modelled as a Gaussian with parameters that were fine-tuned to match that of the slowly rotating solar analogue, 18 Sco. 
The local Stokes V profile for each element across the stellar disc is modelled as the derivative of the local Stokes I profile.  

We model the disc-integrated Stokes I and V profiles from the fully resolved simulated surface field maps using the rotation periods of the simulated stars\footnote{See \citet[section 2.2]{Lehmann2018} for the determination of the rotation periods.} and the corresponding $v_e \sin i$, see Table~\ref{Tab:ProtStars}. As described above, we use a local profile, and assume a fixed number of 30 velocity bins, ranging from $-20\,\rm{km/s}$ to $20\,\rm{km/s}$. Each time-series consists of 25 sets of Stokes~IV profiles corresponding to 25 equally spaced observational phases. As explained earlier, the brightness distribution cannot be reconstructed for slowly rotating stars with activity levels similar to our models, so we assume uniform brightness when generating the disc-integrated Stokes I and V profiles.

We add Gaussian noise to our simulated profiles. Published maps of similarly active stars are based on noise levels of between 3-5\% of the maximum amplitude of the Stokes V signature, which is in line with several observational datasets, e.g., $\epsilon$\,Eri \citep{Jeffers2014}, 61\,Cyg\,A \citep{BoroSaikia2016}. As the amplitudes of the Stokes V signatures can vary by well over an order of magnitude in our simulated datasets, keeping the SNR constant across the simulated sample would result in a much larger relative noise level in the least active models compared to the most active models. We therefore decided to inject a noise level corresponding to $3\,\%$ of the maximum signature in each of our datasets. This corresponds to a continuum SNR of $3\,560\,000$ and $100\,000$ in our least and most active sets of models, respectively -- we note that these are particularly high SNR as they are based  on simulations. This leads to the time-series for even our most active model having an absolute SNR that is $\sim$25\% higher than that typically found in  datasets of solar-activity stars that are used for ZDI. A full list of the average SNR for each stellar model (including inclination angle) is displayed in Table~\ref{Tab:ProtStars}.
 
%
%
%
 \begin{table}
 \caption{The rotation periods, $v_e \sin i$ and averaged SNR for the analysed stars for both inclinations. The stars are characterised by their flux emergences rate (ER) and their differential rotation (DR) in solar terms.}
 \begin{tabular}{c|c|r|r|r|r|r|r}
 \hline
ER & DR & $P_{\mrm{rot}}$& $v_e \sin i$ & $v_e \sin i$ & SNR & SNR\\
$\mrm{[ER_{\odot}]}$ & $\mrm{[DR_{\odot}]}$ & $\mrm{[d]}$& $\mrm{[km/s]}$ & $\mrm{[km/s]}$ &  & \\
& & & $i = 20^{\circ}$ & $i = 60^{\circ}$ & $i = 20^{\circ}$ & $i = 60^{\circ}$\\
 \hline
1 & 1 & 27.00 & 0.64 & 1.62 & 1\,550\,000 & 550\,000\\
3 & 3 & 19.00 & 0.91& 2.31 & 370\,000 & 340\,000\\
5 & 5 & 17.00 &1.02 & 2.58 & 160\,000 & 200\,000\\
 \hline
 \end{tabular}
 \label{Tab:ProtStars}
 \end{table}

\subsection{The effect of $v_e \sin i$ on the resolution of the Stokes~V profiles}
\label{sec:vsinistokes}

Before applying ZDI to our simulated Stokes V profiles, we investigate the impact of the $v_e \sin i$ and of the spatial resolution of the input map on the Stokes~V profiles.
The decomposition of the highly resolved simulated magnetic field maps provides a unique opportunity to analyse the influence of magnetic structures of different length scales on the Stokes~V profiles. We want to estimate at which length scale magnetic structures no longer contribute to the observed Stokes~V profiles by modelling the Stokes~V profiles of magnetic maps including an increasing number of $\ellsum\ $-modes, evaluating the length-scale beyond which the Stokes~V profiles do not change significantly. The Stokes~V profiles are therefore ``blind'' to magnetic structures smaller than the  length scale related to this threshold $\ell$-mode. This is highly dependent on the $v_e \sin i$ of the  star, as the $v_e \sin i$ has a direct impact on the amount of spatial information that can be resolved  with Doppler Imaging techniques.

%
%
%
\begin{figure*}
\centering
\includegraphics[angle=0,width = \textwidth ,clip]{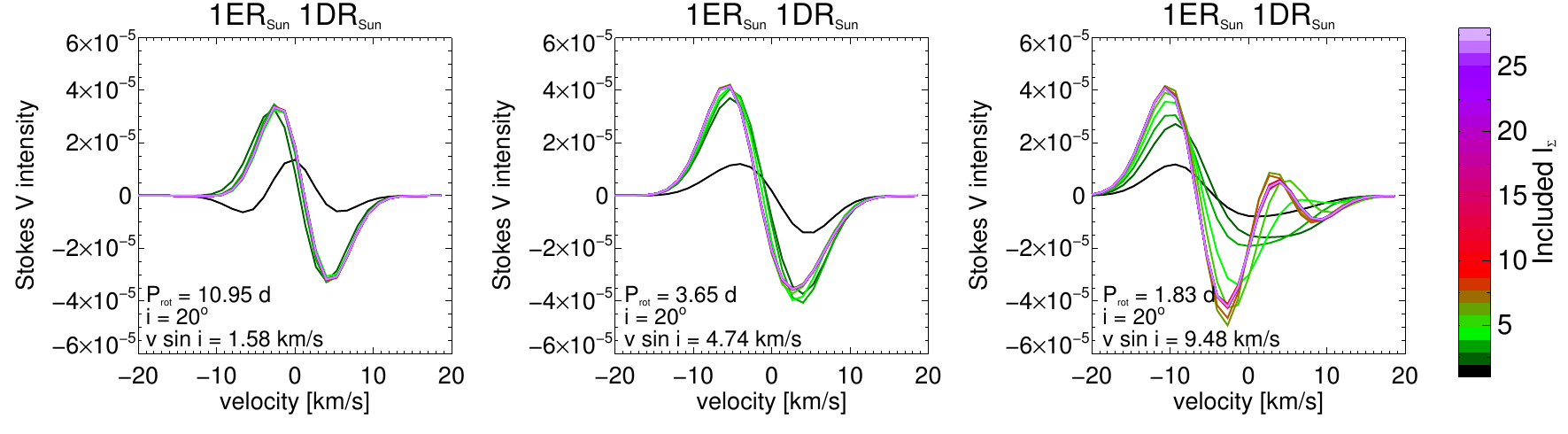}
\includegraphics[angle=0,width = \textwidth ,clip]{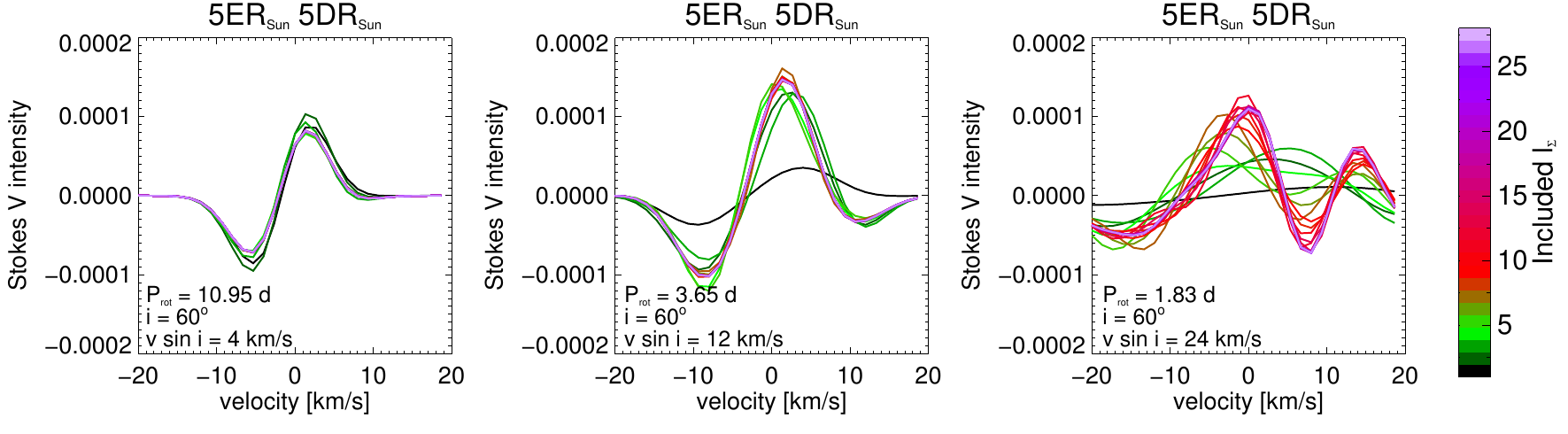}
\caption{The Stokes~V profiles of an example phase modelled from a map of the solar-like star observed at $i=20^{\circ}$(\textit{top row}) and from a map of the most active star ($\mrm{ER} = 5\,$\ERSun\ , $\mrm{DR} = 5\,$\DRSun\ ) observed at $i=60^{\circ}$ (\textit{bottom row}). These are representative of the changes seen over the entire time-series in each case. The rotation periods are artificially increased (from \textit{left} to \textit{right}) and the Stokes~V profiles are generated by including more and more $\ellsum\ $-modes, i.e. smaller scale structures, of the input map (see colour bar on the left).  }
\label{Fig:VsiniStokesV}
\end{figure*}

%
%
%
\begin{figure*}
\centering
\includegraphics[angle=0,width = \textwidth, trim={0 0 0 5cm} ,clip]{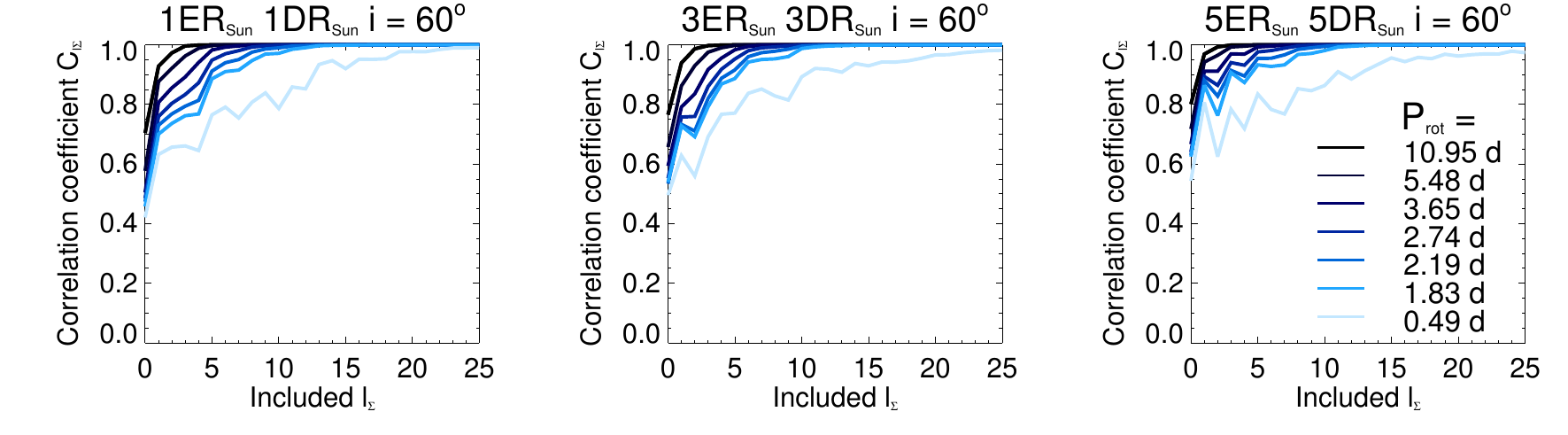}
\caption{The correlation coefficient $C_{\ellsum\ }$ between two successive $\ellsum\ $-modes for the three different stars (\textit{left} to \textit{right}) and the higher inclination $i=60^{\circ}$. The colour of the curves indicates the applied rotation period. }
\label{Fig:VsiniCorrCoeff}
\end{figure*}

We model the Stokes~V profiles for the three stellar activity models, using 10 maps selected over a timeframe of a year, with the input maps including modes $\ellsum\ = 1, 2, \ldots, 28$. We spin up the stars to artificially high rotation periods of $P_{\rm{rot}} \approx 11$ to $\approx 0.5$ days, which leads to values of $v_e \sin i = 4 - 90\,\rm{km/s}, i=60^{\circ}$, see Table~\ref{Tab:vsini}. We also investigate the effects of inclination by modelling profiles assuming inclination angles of $60^{\circ}$ and $20^{\circ}$, see Table~\ref{Tab:vsini}.
Fig.~\ref{Fig:VsiniStokesV} shows the Stokes~V profiles for three different $v_e \sin i$ for a solar map with an inclination of $20^{\circ}$ (\textit{top row}) and for the most active star ($\mrm{ER} = 5\,$\ERSun\ \ and $\mrm{DR} = 5\,$\DRSun\ ) with an inclination of $60^{\circ}$ (\textit{bottom row}). The colour of the Stokes~V profiles indicates which $\ellsum\ $-modes are included (see colour bar on the right). We note that for the low $v_e \sin i \le 5\,\rm{km/s}$, the Stokes~V profiles does not change when including  modes higher than $\ellsum\ = 5$. With increasing $v_e \sin i$, the threshold $\ellsum\ $ also increases, e.g. $\ellsum\ = 15$ for $v_e \sin i = 24\,\rm{km/s}$. Furthermore, the inclination influences the structure of the Stokes~V profiles, as for low inclination only the polar magnetic field region (which tends to look simpler) is visible, while for high inclinations, the more structured  equatorial  region can be detected. 

To quantify the influence of the included $\ellsum\ $-modes, we determine the correlation coefficient between two Stokes~V profiles ($V_{\ellsum\ }$) of successive $\ellsum\ $-modes and averaged them over the ten maps per star ($n_{\rm{map}} = 10$),
\begin{align}
C_{\ellsum\ , \rm{map}} &= \mathtt{correlate}\left(V_{\ellsum\ , \rm{map}}, V_{\ellsum\ + 1, \rm{map}}\right), \\
C_{\ellsum\ } &= \frac{ \sum_{\rm{Map} = 1}^{n_{\rm{map}}} C_{\ellsum\ , \rm{map}} }{ n_{\rm{map}} }.
\end{align}
We calculate $C_{\ellsum\ }$ for several $P_{\rm{rot}}$ (see Table~\ref{Tab:vsini}) and plot $C_{\ellsum\ }$ against $\ellsum\ $ for each star and the higher inclination angle $i=60^{\circ}$ in Fig.~\ref{Fig:VsiniCorrCoeff}. The results for the lower inclination angle $i=60^{\circ}$ can be found in the appendix, see Fig.~\ref{Fig:VsiniCorrCoeff_20}. The colour becomes lighter with decreasing $P_{\rm{rot}}$ (i.e. more rapid rotation). A correlation coefficient of $C_{\ellsum\ } = 1$ indicates that the Stokes~V profiles of that mode $\ellsum\ $ and the following mode $\ellsum\ +1$ are identical, so that the Stokes~V profiles are blind for magnetic structures of the corresponding length scale and smaller.
The correlation coefficients are higher for $i =20^{\circ}$ compared to the values for $i =60^{\circ}$ due to the lower $v_e \sin i$ and the less complex structures at the poles. An observer can only observe magnetic structures for $\ell \leq 5-7$ which correspond to an angular resolution of $\theta \gtrapprox 36^{\circ}-25.7^{\circ}$ for low inclination angle of $20^{\circ}$ and $P_{\rm{rot}} \geq 1.8\,\rm{days}$. Also for a higher inclination of $60^{\circ}$ an observer can only resolve structures $\ell \leq 5-7$ for the slow rotators with $P_{\rm{rot}} \geq 3.6\,\rm{days}$. For the faster rotators with $P_{\rm{rot}} \leq 1.8\,\rm{days}$ we can resolve structures down to $\ell = 15$ corresponding to an angular resolution of $\theta \approx 12^{\circ}$ and for the very fast rotators structures down to a few degree sizes. The correlation coefficients for the most active star ($\mrm{ER} = 5\,$\ERSun\ , $\mrm{DR} = 5\,$\DRSun\ ), $i =60^{\circ}$, show a zig-zag pattern for the lower rotation periods, see Fig.~\ref{Fig:VsiniCorrCoeff} \textit{right}. The large-scale magnetic field structure of this star is dominated by two toroidal bands of opposite polarity. That signature is best captured by the even $\ell$-modes. The inclusion of even $\ellsum\ $ modes causes a larger difference to the Stokes~V profiles (lower correlation coefficients) than the inclusion of additional odd $\ellsum\ $.

%
%
%
 \begin{table}
 \caption{The artificially increased rotation periods and the corresponding velocities $v_e$ and $v_e \sin i$ for investigating the effect of $v_e \sin i$ on the resolution of the Stokes~V profiles.}
 \centering
 \begin{tabular}{r|r|r|r}
 \hline
$P_{\mrm{rot}}$& $v_e$  & $v_e \sin i$ & $v_e \sin i$ \\
$\mrm{[d]}$& $\mrm{[km/s]}$  & $\mrm{[km/s]}$ & $\mrm{[km/s]}$ \\
  &  & $i = 20^{\circ}$ & $i = 60^{\circ}$  \\
 \hline
 10.95 & 4.62 & 1.58 & 4.00 \\
 5.48 & 9.24 & 3.16 & 8.00 \\
 3.65 & 13.86 & 4.74 & 12.00 \\
 2.74 &18.48 & 6.32 & 16.00 \\
 2.19 & 23.10 & 7.90 & 20.00 \\
 1.83 & 27.70 & 9.48 & 24.00 \\
 0.49 & 104.00& 35.54 & 90.00 \\
 \hline
 \end{tabular}
 \label{Tab:vsini}
 \end{table}

\subsection{Fitting the Stokes V profiles}

\begin{landscape}
%
%
%
\begin{figure}
\centering
\includegraphics[angle=0,width = 0.21\textwidth ,clip]{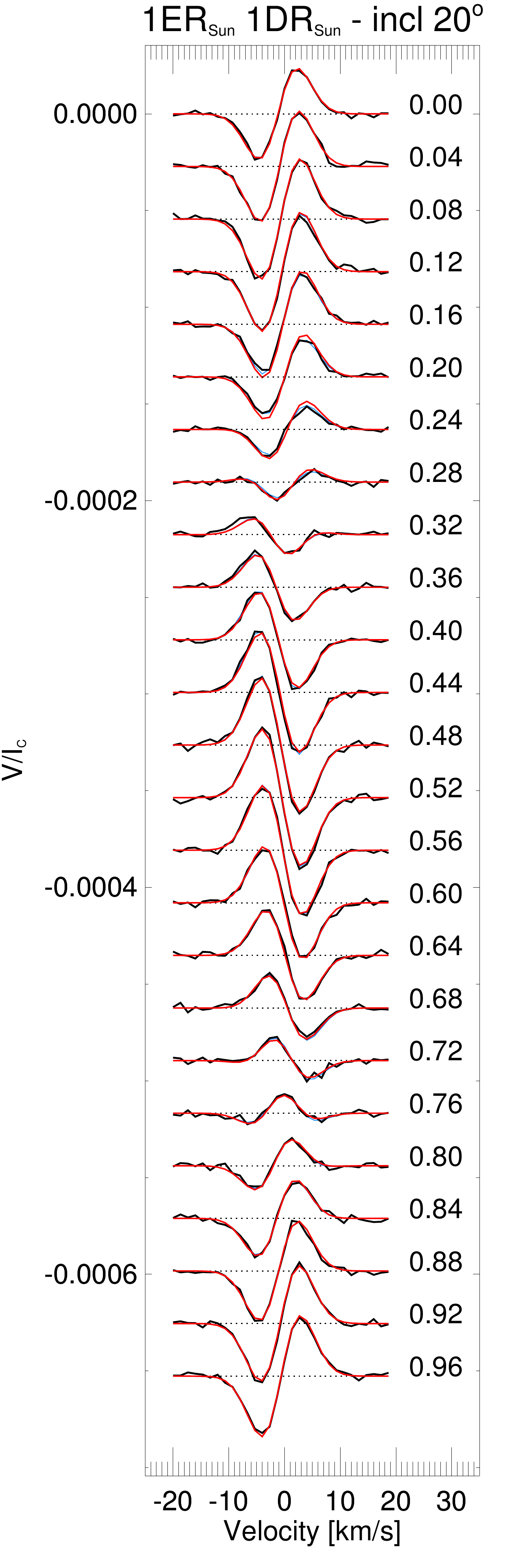} 
\includegraphics[angle=0,width = 0.21\textwidth ,clip]{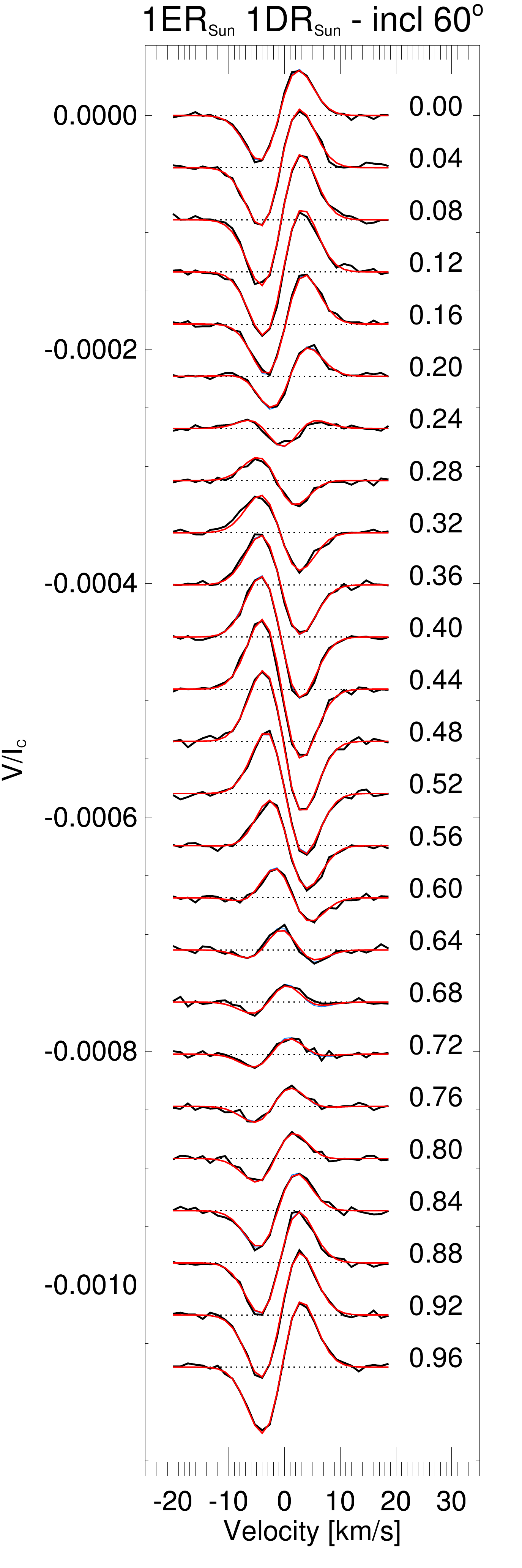} 
\hspace{0.25cm}
\includegraphics[angle=0,width = 0.21\textwidth ,clip]{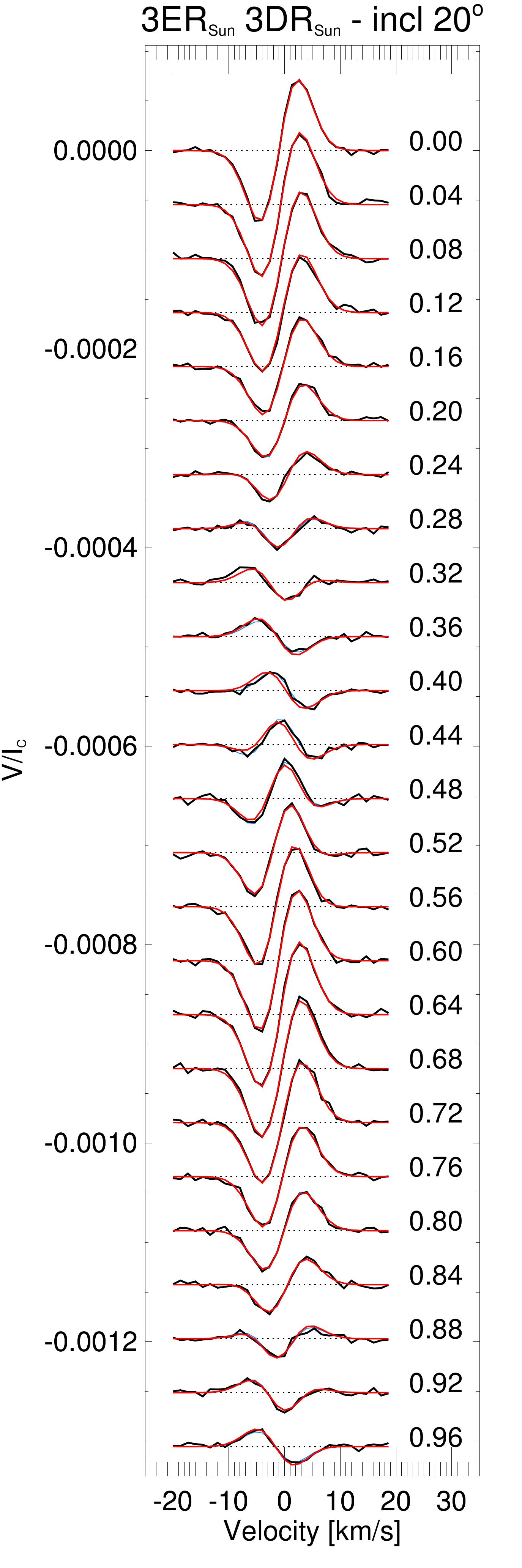} 
\includegraphics[angle=0,width = 0.21\textwidth ,clip]{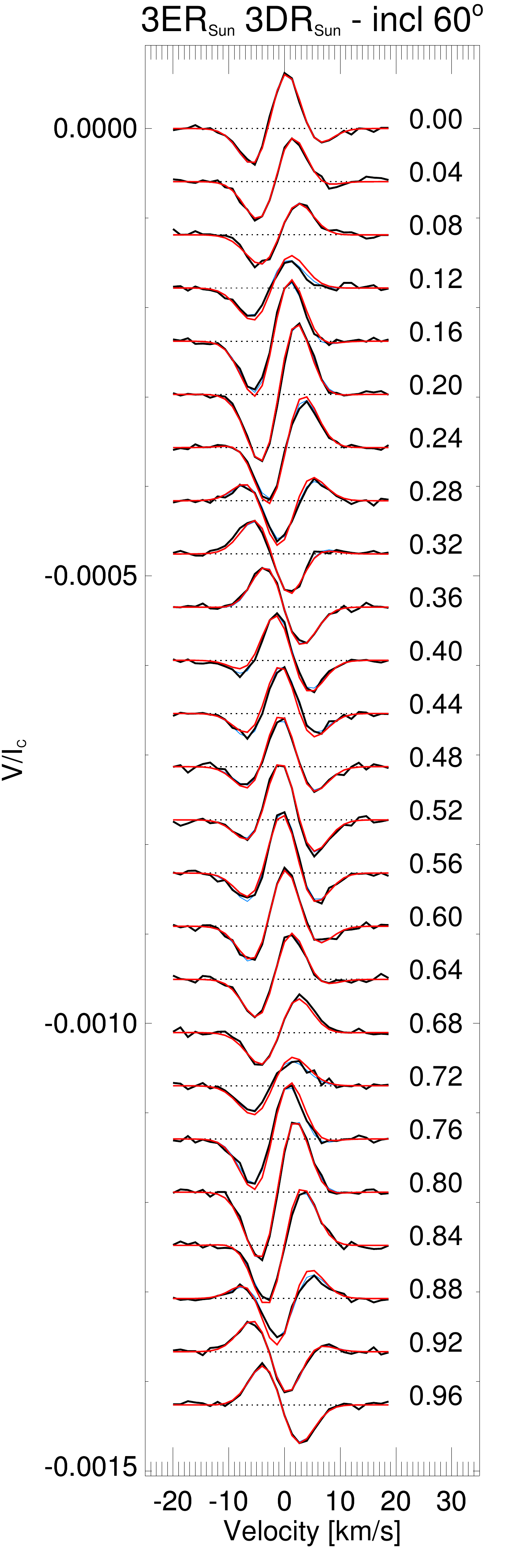} 
\hspace{0.25cm}
\includegraphics[angle=0,width = 0.21\textwidth ,clip]{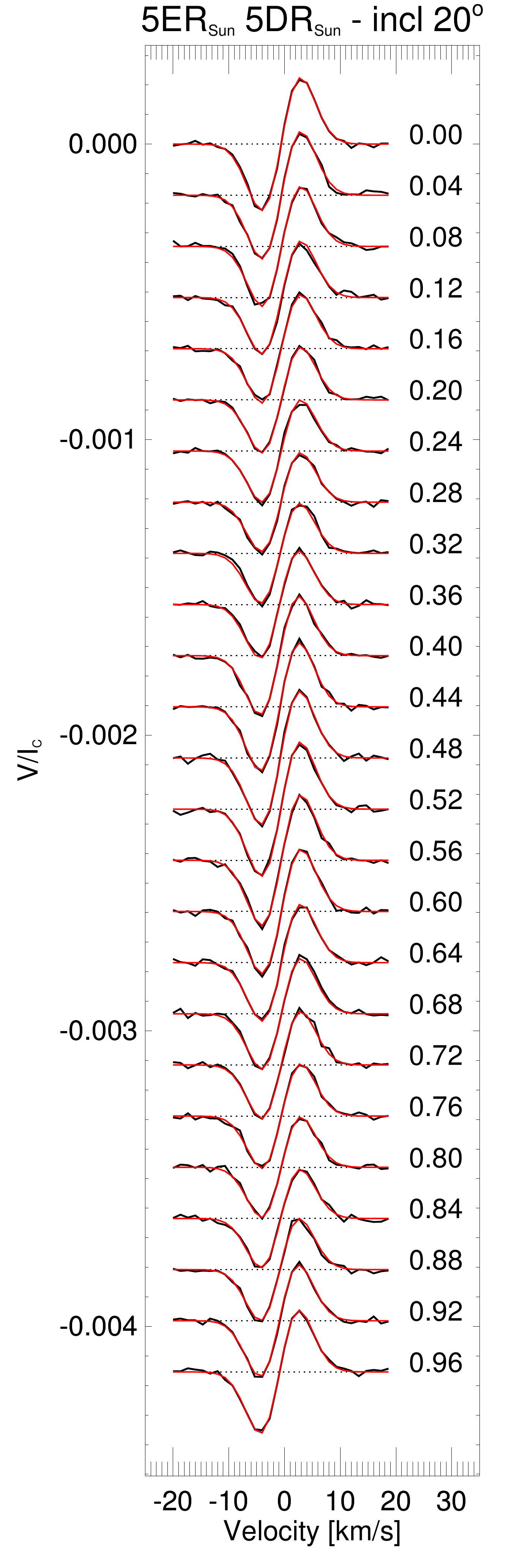} 
\includegraphics[angle=0,width = 0.21\textwidth ,clip]{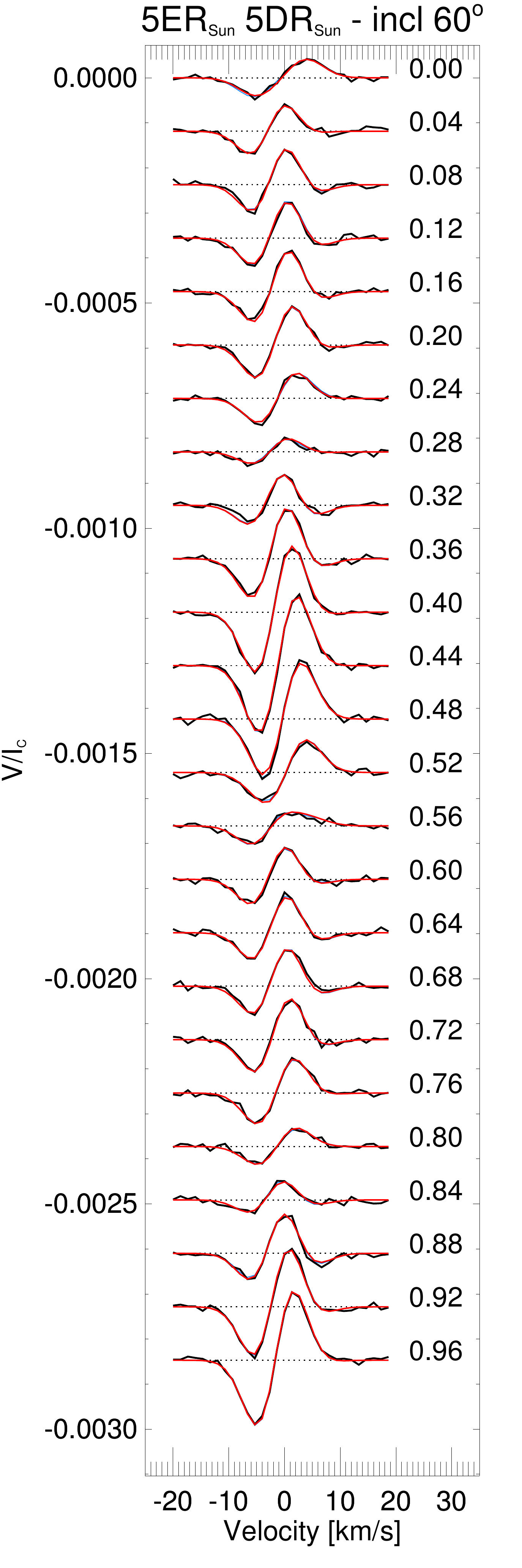} 

\caption{Six example sets of Stokes~V profiles (one per inclination and stellar model) showing the fit of the Stokes~V profiles modelled from the fully resolved input map with ZDI. In each case, the thick solid black line is the noisy Stokes~V profile, that is fitted with ZDI. The red line is the resulting ZDI fit and the blue line (mostly behind the red line) is the noise-free Stokes~V profile of the fully resolved input map. The dotted black line is the null-line. The phases are written to the right.
}
\label{Fig:StokesIVFit}
\end{figure}

\end{landscape}

We then proceed with applying ZDI to our simulated Stokes~V profiles and try to find the best agreement with these input datasets.
Fig.~\ref{Fig:StokesIVFit} presents our ZDI fits to the Stokes~V profiles for one example map per stellar activity model and inclination. The black lines represent the noisy Stokes V profiles generated using the fully resolved simulated maps and the red lines are our best fit using ZDI. The blue lines (mostly hidden behind the red lines of the ZDI fits) show the noise-free Stokes V profiles for comparison. The observational phases are written to the right of the single profiles. ZDI fits the noisy Stokes V profiles very well. On average we achieve a reduced $\chi_r^2 \approx 3$. Our best fit reaches a $\chi_r^2 = 1.05$ and only four out of 60 maps had a $\chi_r^2 \ge 5$. Our degree of fit is comparable to the degree of fit for observations, e.g. for the solar-like stars $\varepsilon$~Eri and HN~Peg \citep{Jeffers2014,BoroSaikia2015}. In most cases the ZDI reconstructed profiles (red lines) fit the noise-free simulated Stokes~V profiles (blue lines) although we apply the ZDI fitting onto the noisy Stokes~V profiles (black line). The two different inclinations show no significant effect on the degree of fitting.

\section{Results}
\label{Sec:Results}

In this section we analyse of the ability of the ZDI technique to reproduce the input 3D non-potential magnetic field simulations of solar-like stars.

\subsection{Comparing the input maps with the ZDI reconstructions}
\label{SubSec:ComparingTheInputMapsWithTheZDIMaps}

\begin{landscape}

%
%
%
\begin{figure}
\raggedright
\large
\hspace{0.7cm}\textbf{Input simulation}\hspace{2.3cm}\textbf{Large-scale field of the input simulation}\hspace{4.2cm}\textbf{ZDI reconstruction}\\
\hspace{0.9cm}(Full resolution) \hspace{3.1cm} ($\ellsum\ = 7$) \hspace{2.6cm} ($\ellsum\ = 5$) \hspace{3.8cm} ($i = 60^{\circ}$) \hspace{2.6cm} ($i = 20^{\circ}$)\\
\centering
\includegraphics[scale = 0.13]{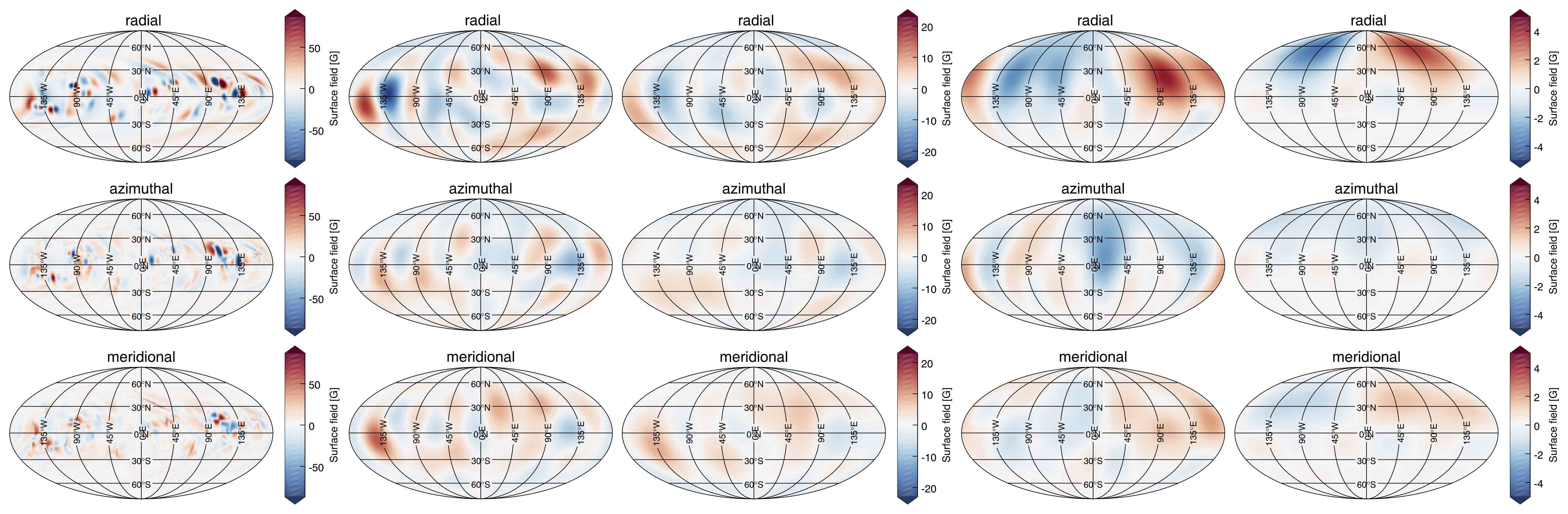}
\caption{The Mollweide projected magnetic field maps for one example of the solar-like star ($\mrm{ER} = 1\,$\ERSun\ \ and $\mrm{DR} = 1\,$\DRSun\  ). 
\textit{From left to right}: the simulated fully resolved simulated input map, the large-scale field of the simulated input map for $\ell_{\Sigma} = 7$ and $5$ compared with the ZDI reconstructed maps applying an inclination of $60^{\circ}$ and $20^{\circ}$. The radial component is displayed in the \textit{top row}, the azimuthal in the \textit{middle row} and the meridional component in the \textit{bottom row}.}
\label{Fig:11_1957_ZDI}
\end{figure}

\end{landscape}

\begin{landscape}

%
%
%
\begin{figure}
\raggedright
\large
\textbf{a.}\hspace{0.4cm}\textbf{Input simulation}\hspace{2.3cm}\textbf{Large-scale field of the input simulation}\hspace{4.2cm}\textbf{ZDI reconstruction}\\
\hspace{0.9cm}(Full resolution) \hspace{3.1cm} ($\ellsum\ = 7$) \hspace{2.6cm} ($\ellsum\ = 5$) \hspace{3.8cm} ($i = 60^{\circ}$) \hspace{2.6cm} ($i = 20^{\circ}$)\\
\centering
\includegraphics[scale = 0.13]{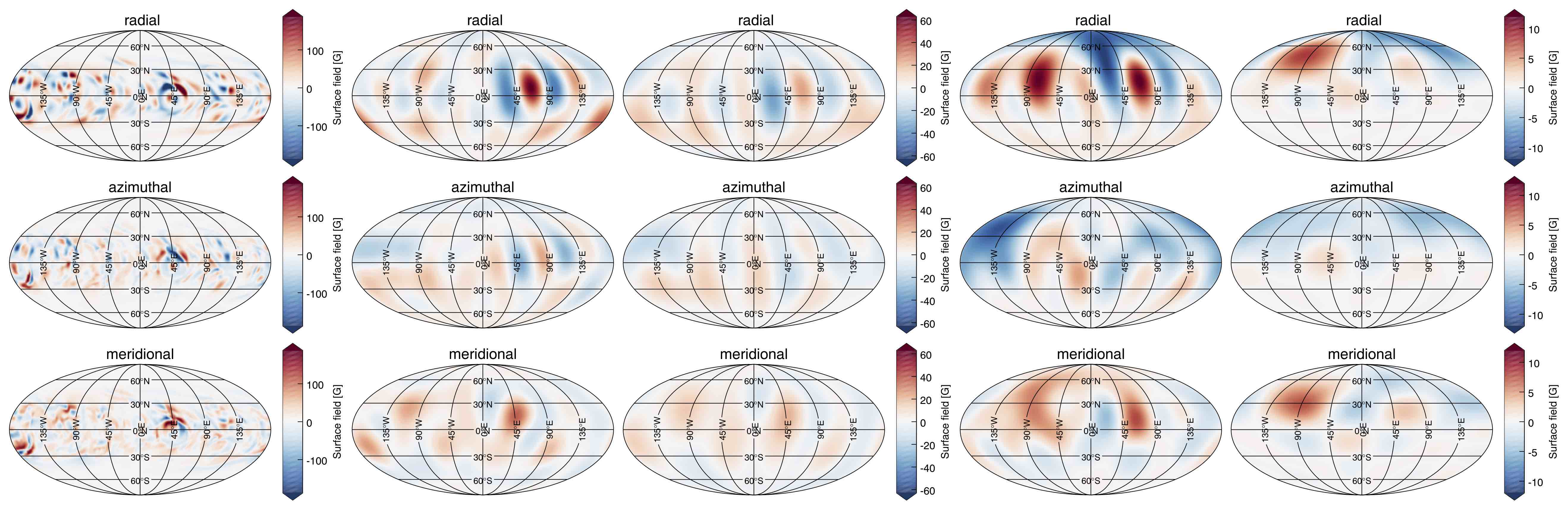}\\
\raggedright
\large
\textbf{b.}\hspace{0.4cm}\textbf{Input simulation}\hspace{2.3cm}\textbf{Large-scale field of the input simulation}\hspace{4.2cm}\textbf{ZDI reconstruction}\\
\hspace{0.9cm}(Full resolution) \hspace{3.1cm} ($\ellsum\ = 7$) \hspace{2.6cm} ($\ellsum\ = 5$) \hspace{3.8cm} ($i = 60^{\circ}$) \hspace{2.6cm} ($i = 20^{\circ}$)\\
\includegraphics[scale = 0.13]{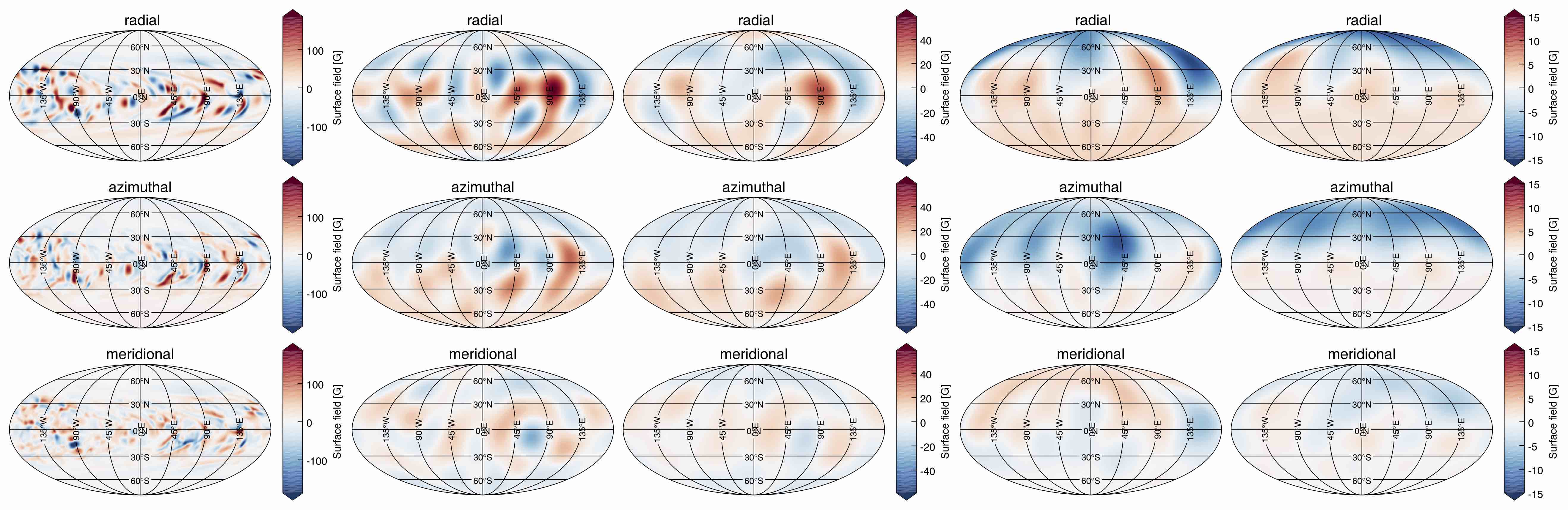}
\caption{The Mollweide projected magnetic field maps for \textbf{a.} one example of the more active star ($\mrm{ER} = 3\,$\ERSun\ \ and $\mrm{DR} = 3\,$\DRSun\  ) and \textbf{b.} for one example of the most active star ($\mrm{ER} = 5\,$\ERSun\ \ and $\mrm{DR} = 5\,$\DRSun\  ). 
The format is the same as for Fig.~\ref{Fig:11_1957_ZDI}.}
\label{Fig:33_1992_ZDI}
\end{figure}

\end{landscape}

At first, we compare the input maps and the large-scale field of the input maps with the reconstructed ZDI maps. One example map per stellar model and inclination is plotted in the Figs.~\ref{Fig:11_1957_ZDI}-\ref{Fig:33_1992_ZDI}b. These figures show Mollweide projected maps for the radial (\textit{top row}), azimuthal (\textit{middle row}) and meridional field components (\textit{bottom row}) for one of the ten maps for the solar-like star ($\mrm{ER} = 1\,$\ERSun\ , $\mrm{DR} = 1\,$\DRSun\ , Fig.~\ref{Fig:11_1957_ZDI}), the more active star ($\mrm{ER} = 3\,$\ERSun\ , $\mrm{DR} = 3\,$\DRSun\ , Fig.~\ref{Fig:33_1992_ZDI}a) and the most active star ($\mrm{ER} = 5\,$\ERSun\ , $\mrm{DR} = 5\,$\DRSun\ , Fig.~\ref{Fig:33_1992_ZDI}b). Figs.~\ref{Fig:11_1957_ZDI}-\ref{Fig:33_1992_ZDI}b present the simulated input map from which the ZDI fitted Stokes~V profiles are generated in the \textit{first column}. The \textit{second and third columns} show the large-scale field of the simulated input map restricted to $\ellsum\ = 7$ and 5, respectively. The two last columns display the  reconstructed ZDI maps for an inclination $i = 60^{\circ}$ (\textit{4th column}) and for an inclination $i = 20^{\circ}$ (\textit{5th column}). Note that the maps are plotted with three different colour bars for the simulated input map, the large-scale field maps and the ZDI reconstructed maps. It is more instructive to focus on a comparison of the morphology and not on the colourscale, as clearly the reconstructions recover much less field than is originally input. The ZDI reconstructed maps presented in Figs.~\ref{Fig:11_1957_ZDI}-\ref{Fig:33_1992_ZDI}b correspond to the Stokes~V profile fits shown in Fig.~\ref{Fig:StokesIVFit}; these represent typical fits for the lower and higher activity models, and the best fits for the moderate activity model.

In general, the ZDI reconstructed maps look very different from the simulated input maps. In the simulated input maps (Figs.~\ref{Fig:11_1957_ZDI}-\ref{Fig:33_1992_ZDI}b, \textit{1st column})  the small-scale field structures originating from the active regions dominate, whereas the ZDI reconstructed maps (Figs.~\ref{Fig:11_1957_ZDI}-\ref{Fig:33_1992_ZDI}b, \textit{4th and 5th column}) only show large-scale field. As shown in Section~\ref{sec:vsinistokes}, the Stokes~V profiles only provide  information about the large-scale field especially in slowly rotating stars, and miss the small-scale field. It is therefore advisable to compare the ZDI reconstructed maps with the large-scale field of the input maps, (Figs.~\ref{Fig:11_1957_ZDI}-\ref{Fig:33_1992_ZDI}b, \textit{2nd and 3rd column}). We allowed the ZDI code to use $\ell$-modes up to $\ell =7$ but find that most of the information in the reconstructions is concentrated in the $\ell$-modes $\ell = 1-5$, so we present comparisons with the input large-scale field maps restricted to both $\ellsum\ = 7$ and $5$. ZDI consistently recovers significantly less magnetic field than the large-scale field of the input maps, which is likely due to flux cancellation in the Stokes V profiles. Here, we compare the magnetic field values averaged over the whole surface, not only the maximum visible surface of the input maps (i.e. we do not account for inclination effects). Accounting for inclination would reduce the magnetic field values of the input map by a few percent.

The ZDI reconstructed maps for the solar-like star ($\mrm{ER} = 1\,$\ERSun\ , $\mrm{DR} = 1\,$\DRSun\ , Fig.~\ref{Fig:11_1957_ZDI}) show significantly different morphologies for the two different inclinations, with the low inclination reconstructions showing much less structure in all three vector components. The ZDI reconstructions at higher inclination ($i=60^{\circ}$) show a much better agreement to the large-scale field of the input than the reconstruction for the lower inclination $i=20^{\circ}$, with the main features of the large-scale field $\ellsum\ = 5$ of the northern hemisphere recovered down to the equatorial region. The reconstructed azimuthal field also recovers most of the main features of the northern hemisphere, but with a higher uncertainty. The meridional field is affected by crosstalk with the radial field and shows the least agreement as expected \citep{Donati1997}. The low inclination maps are restricted to the polar view and a worse spatial resolution as the inclination leads the $v_e \sin i$ to be reduced from $v_e \sin i = 1.62\,\mrm{km/s}$ (for $i = 60^{\circ}$) to $v_e \sin i = 0.64\,\mrm{km/s}$. With these very low $v_e \sin i$ values we are limited by the thermal width of the line profile as well as the spectral resolution of the instrumentation.

The ZDI reconstructed maps for the more active star ($\mrm{ER} = 3\,$\ERSun\ , $\mrm{DR} = 3\,$\DRSun\ , Fig.~\ref{Fig:33_1992_ZDI}a) show a higher agreement to the large-scale field of the input compared to the solar-like star. In particular, the radial map for the higher inclination star shows remarkably good agreement with the large-scale field $\ellsum\ = 5$, Fig.~\ref{Fig:33_1992_ZDI}a (\textit{top row, 3rd and 4th column}). All features of the large-scale field are reconstructed by ZDI down to south of the equatorial region. Remember that two different colour scales are applied by comparing the two maps. The main features of the azimuthal and meridional field maps are also recovered, though less accurately than in the radial map. The meridional vector is less affected by crosstalk compared to the solar-like star map. As there is an increased contribution in the input meridional field map at this activity level compared to the solar-like star, there is a stronger contribution from this component to the Stokes V profile. The ZDI reconstruction at lower inclination (Fig.~\ref{Fig:33_1992_ZDI}a, \textit{5th column}) displays worse spatial resolution, and is more restricted to the structure in  the northern hemisphere; there is still an acceptable agreement with the large-scale field in the input map.

The reconstructed ZDI maps for the most active star ($\mrm{ER} = 5\,$\ERSun\ , $\mrm{DR} = 5\,$\DRSun\ , Fig.~\ref{Fig:33_1992_ZDI}b) generally show good agreement with the large-scale field of the input maps, Fig.~\ref{Fig:33_1992_ZDI}a shows a particularly good example of this. The ZDI reconstruction for the higher inclination shows poorer spatial resolution than expected from the large-scale field $\ellsum\ = 5$ of the input maps. The ZDI reconstruction at lower inclination shows less structure. It is particularly noticeable that the azimuthal field is very well constrained. It becomes dominant and displays a strong belt of negative polarity on the northern hemisphere. This azimuthal field gains more structure as the inclination increases. This strong ring also dominates the Stokes~V profiles, especially at low inclination, see Fig.~\ref{Fig:StokesIVFit}, \textit{5th and 6th panel}. The original input map (Fig.~\ref{Fig:33_1992_ZDI}b, \textit{1st column}) of this active star shows many active regions, which have a predominantly negative polarity in the azimuthal field in the northern hemisphere, and a predominantly positive polarity on the southern hemisphere. The global properties of this flux emergence pattern (two azimuthal bands of opposite polarity at mid latitudes) shape the large-scale field $\ellsum\ = 5$ and are therefore dominant in the ZDI reconstruction. In particular for the low inclination where the star is seen nearly pole-on, ZDI is most sensitive to magnetic features, as the polar region contains relatively low levels of magnetic flux.

To summarise: ZDI recovers the visible structure of the large-scale field morphology of solar-like stars but loses the magnetic field strength. Limiting the original input maps to the large-scale field via the spherical harmonic decomposition (see \citealt{Lehmann2017, Lehmann2018}) provides a very good approximation of the magnetic field morphology structure that will be recovered by ZDI. Although ZDI is allowed to use $\ell$-modes up $\ell = 7$ it recovers structures up to $\ell \leq 5$ most of the time. All 60 analysed maps show significantly lower magnetic field strengths than the corresponding large-scale field of the input maps, see also Section \ref{subsec:RoveringLargeScaleProperties}. Furthermore, ZDI is affected by the inclination: a lower inclination (pole-on view) reduces the spatial resolution and further restricts the view to one hemisphere. 

%

\subsection{Recovering the large-scale field properties}
\label{subsec:RoveringLargeScaleProperties}

%
%
%
\begin{figure}
\centering
\includegraphics[width = 0.49\textwidth ,clip]{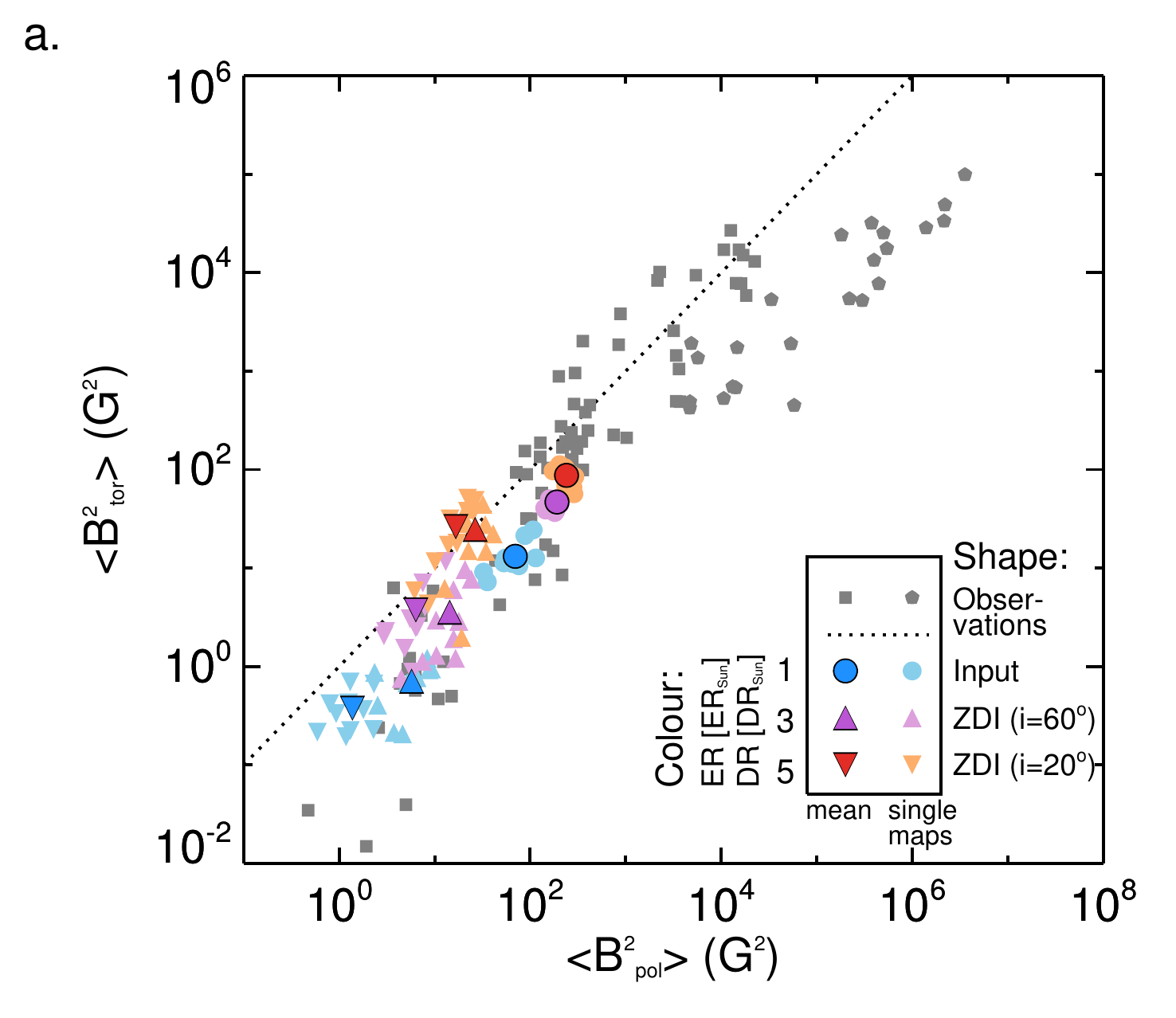} \\
\includegraphics[width = 0.49\textwidth ,clip]{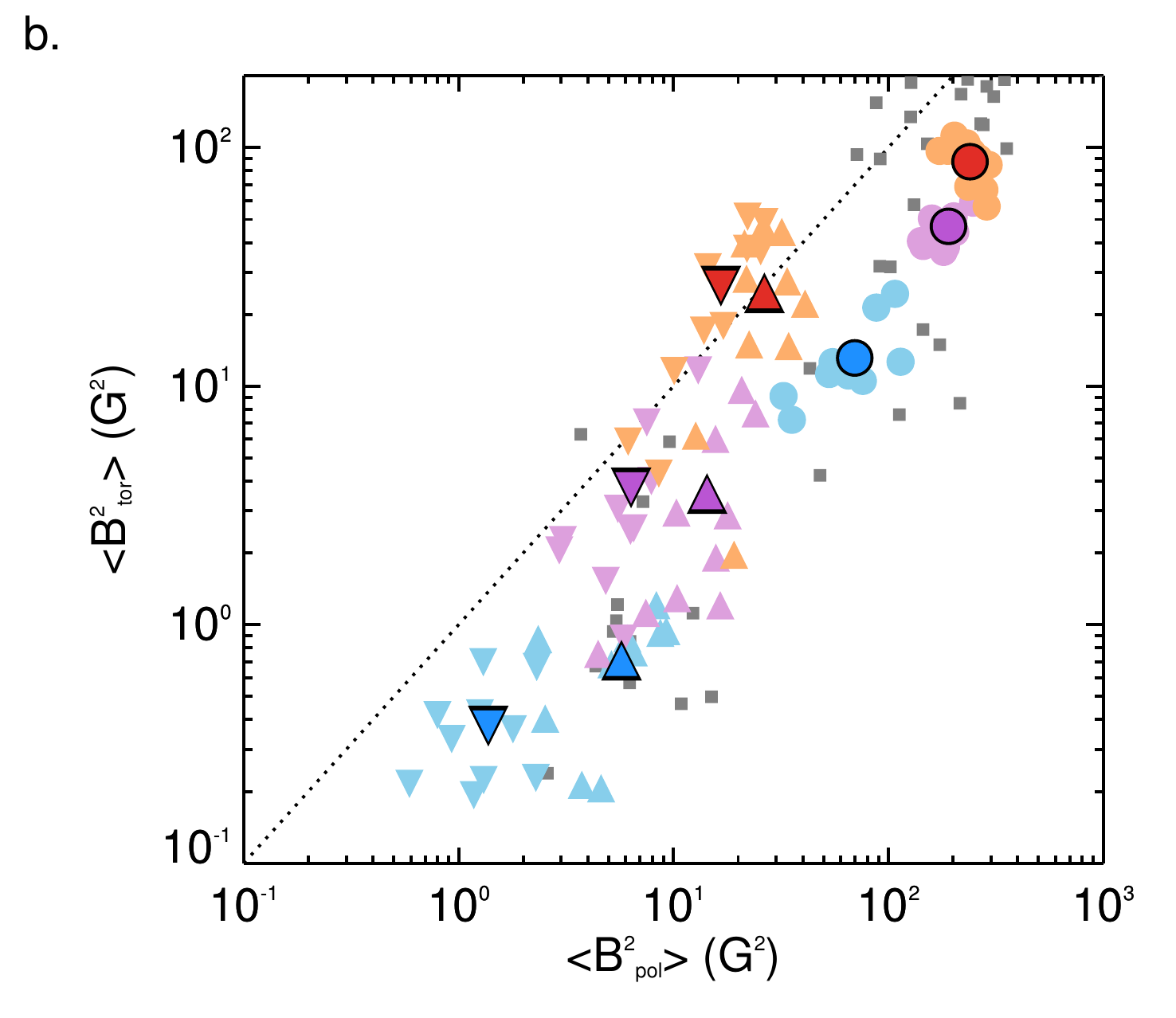} 
\caption{The toroidal against the poloidal magnetic energy for the observed cool stars (grey symboles), the simulated input stars with $\ellsum\ = 7$ (circles) and the ZDI reconstructed maps for inclination $i = 60^{\circ}$ (triangle) and $i = 20^{\circ}$ (upside down triangle). The colour indicates the activity of the star: blue for the solar-like star ($\mrm{ER} = 1\,$\ERSun\ \ and $\mrm{DR} = 1\,$\DRSun\ ), purple for the more active star ($\mrm{ER} = 3\,$\ERSun\ \ and $\mrm{DR} = 3\,$\DRSun\ ) and red for the most active star ($\mrm{ER} = 5\,$\ERSun\ \ and $\mrm{DR} = 5\,$\DRSun\ ). We plot all ten maps per star with a fainter colour and smaller symbol size and the average over the ten maps per star with the bolder colour and larger symbol size. The dashed line indicates the unity line. \textbf{a.} The full parameter range covered by the observations. \textbf{b.} A zoom in to the parameter range covered by the input and reconstructed maps.
}
\label{Fig:EpolEtor}
\end{figure}

By comparing the large-scale field of the input maps with that in the ZDI reconstruction we find that ZDI recovers significantly less magnetic field than input, see Figs.~\ref{Fig:11_1957_ZDI}-\ref{Fig:33_1992_ZDI}b. We analyse this further by computing the mean-squared flux density $\langle B^2\rangle\mathrm{[G^2]}$ of the poloidal and toroidal field for the large-scale field input maps and the corresponding ZDI maps, see Fig.~\ref{Fig:EpolEtor}. The mean-squared flux density is a good proxy for the magnetic energy\footnote{The mean squared flux density $\langle B^2\rangle\mathrm{[G^2]}$ is not exactly equivalent but often referred as magnetic energy, see e.g. the review from \cite{Reiners2012}. Be aware that  $\langle B^2\rangle\mathrm{[G^2]}$ for the reconstructions is restricted to the net magnetic flux of the resolution elements. }. 
Fig.~\ref{Fig:EpolEtor} plots the toroidal against the poloidal energy. The top panel, Fig.~\ref{Fig:EpolEtor}a, shows the full range of energy values, set by published maps of cool stars, and the bottom panel, Fig.~\ref{Fig:EpolEtor}b, provides a detailed view over the energy range set by our models and maps. The grey symbols represent the observed maps of cool stars collected by the Bcool and Toupies surveys, and were published by Petit (in preparation); \cite{BoroSaikia2015, doNascimento2014, Donati2003, Donati2008, Fares2009, Fares2010, Fares2012, Fares2013, Folsom2016, Morin2008a, Morin2008, Morin2010, Jeffers2014,  Petit2008, Waite2011}. 
Stars with masses smaller $0.5\,\mrm{M_{\odot}}$ are plotted as pentagons, stars with masses equal or higher than $0.5\,\mrm{M_{\odot}}$ as squares. The coloured symbols represent our analysed maps. The circular symbols stand for the large-scale field of the input maps, i.e. truncating the simulation map up to mode $\ellsum\ = 7$, and the triangular symbols for their ZDI reconstructed maps, where the normal triangles displays the ZDI reconstructions for an inclination of $i=60^{\circ}$ and the up-side-down triangles the ZDI reconstructions for an inclination of $i=20^{\circ}$. The magnetic energy values of the input maps are calculated for the whole surface not acounting for obscuration due to inclination effects.
The colour indicates the activity level of the star: blue for the maps of the solar-like star, purple for the maps of the more active star with $\mrm{ER} = 3\,$\ERSun\ \ and $\mrm{DR} = 3\,$\DRSun\ \ and red for the maps of the most active star with $\mrm{ER} = 5\,$\ERSun\ \ and $\mrm{DR} = 5\,$\DRSun\ . We present the results of the ten single maps per star (fainter coloured symbols) and their average  (stronger coloured and black bordered symbols), see also legend in Fig.~\ref{Fig:EpolEtor}a. The dashed line indicates equal poloidal and toroidal energies.

The large-scale field of the 3D non-potential flux transport simulations show similar values for the poloidal and toroidal magnetic energy to the published ZDI maps of solar-like stars (see \citealt{Lehmann2018} for the full analysis). The ZDI reconstructed maps (triangles) still lie within the spread of the observations but show approximately one order of magnitude less magnetic energy in both components compared to the large-scale field of the input maps (circles), see Fig.~\ref{Fig:EpolEtor}b. For both high and low inclination angles, ZDI reconstructions seems to follow the power law $\langle B^2_{\mathrm{tor}}\rangle \propto \langle B^2_{\mathrm{pol}}\rangle^{1.25\pm0.06}$ found by \cite{See2015} for stars with masses equal or higher than $0.5\,\rm{M_{\odot}}$. ZDI recovers the trends with activity but with systematically lower magnetic energies. The Fig.~\ref{Fig:EpolEtor}b provides a closer view of these results. We note that the spread of the ten maps per star is larger for the ZDI reconstructions (triangles) than for the input maps (circles) but smaller than the spread in the ZDI maps obtained from observations. The ZDI reconstructions for higher inclinations (equator on, normal triangles) show closer values to the large-scale field of the input maps. Furthermore, we see that with lower inclination, less poloidal energy tends to be reconstructed, and for the more active stars a slightly higher amount of toroidal energy is recovered compared to higher inclinations. 

%
%
%
\begin{figure}
\centering
\includegraphics[width = 0.49\textwidth ,clip]{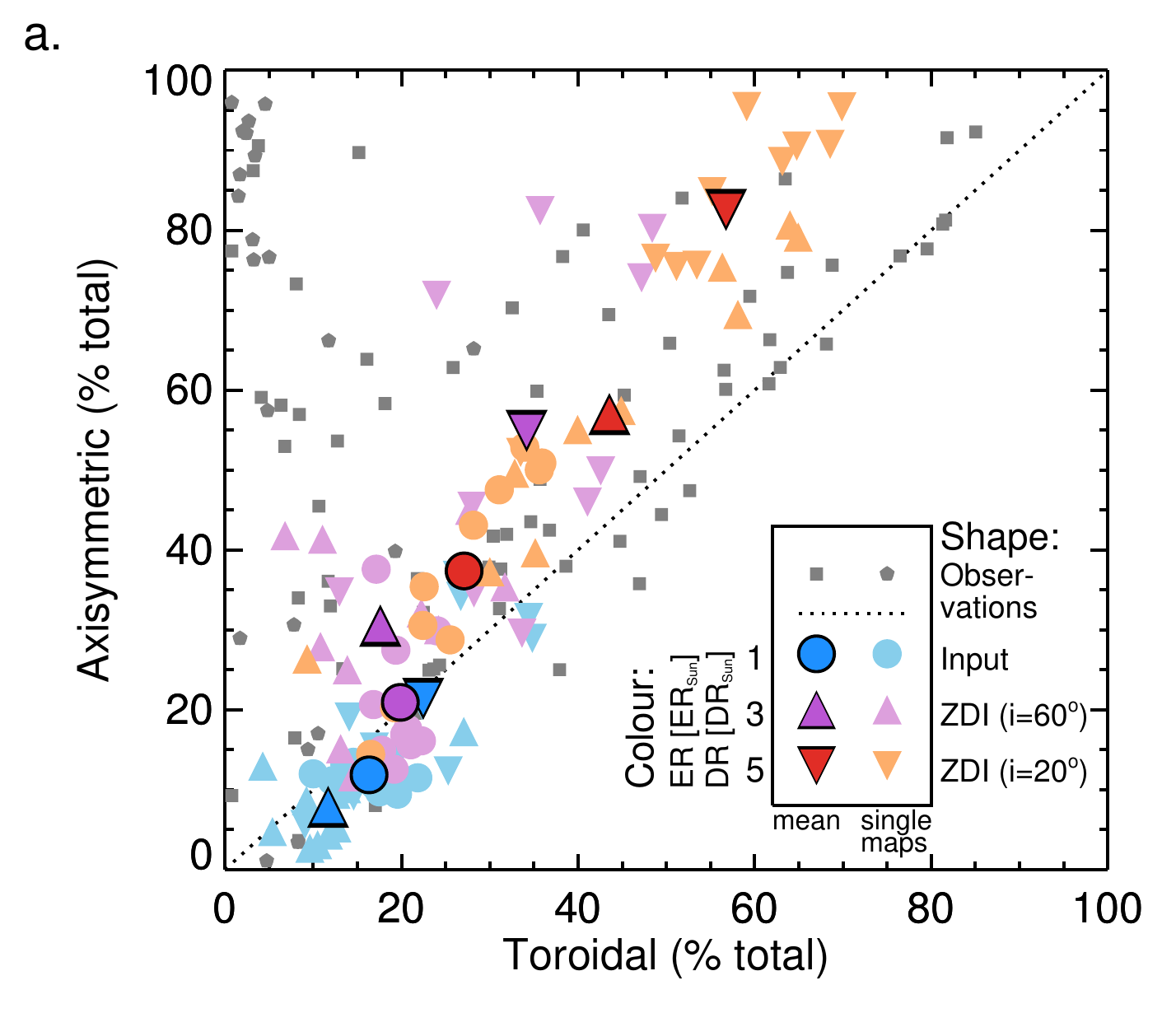} \\
\includegraphics[width = 0.49\textwidth ,clip]{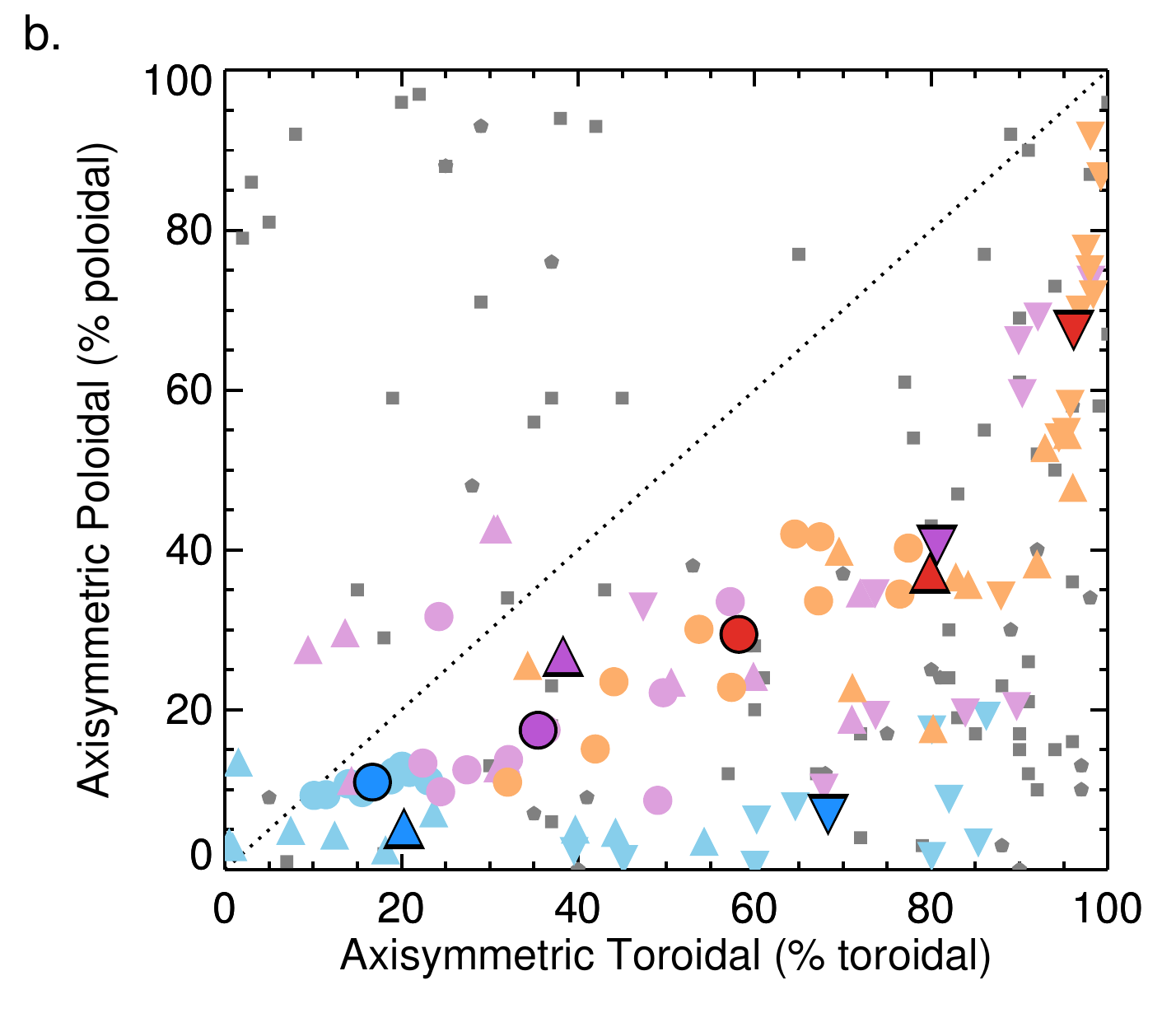} 
\caption{The percentage of axisymmetric fraction against the percentage of toroidal fraction (\textbf{a.}) and the percentage of the axisymmetric poloidal fraction against the percentage of the axisymmetric toroidal fraction (\textbf{b.}) for the observation, large-scale field input maps and ZDI reconstruction presented in Fig.~\ref{Fig:EpolEtor}. The same format as in Fig.~\ref{Fig:EpolEtor} is used.}
\label{Fig:AxiTorAxiPolAxiTor}
\end{figure}

Fig.~\ref{Fig:AxiTorAxiPolAxiTor}a shows the axisymmetric fraction of the field  against the toroidal fraction using the same data and format as in Fig.~\ref{Fig:EpolEtor}. The ZDI reconstructed maps lie again in the same parameter range as the observations. They also follow the trend found by \cite{See2015}: the more toroidal the field morphology, the more axisymmetric it tends to be. For the solar-like star (blue symbols) the fraction of axisymmetric and toroidal field are well recovered within in $\pm20\%$ for both inclinations, but the more active the star the more ZDI tends to overestimate the percentage of axisymmetric and toroidal field. This trend is enhanced by the effect of inclination, with ZDI reconstructing even higher values for toroidal and axisymmetric fields at lower inclination. 

Fig.~\ref{Fig:AxiTorAxiPolAxiTor}b displays the axisymmetric poloidal fraction and the axisymmetric toroidal fraction using the same data and format as in Fig.~\ref{Fig:EpolEtor}. The axisymmetry of the poloidal field is well recovered for the higher inclinations (normal triangles) but the more active the star the more is the axisymmetric toroidal fraction overestimated. For the lower inclination (up-side-down triangles) we see that the more active the star is, the more the axisymmetric poloidal field is overestimated and that the axisymmetric toroidal field is in general 40-50\% higher than excepted. 

The overestimation of the toroidal and axisymmetric fraction especially for the active stars and lower inclinations is likely due to the projection effects of the large-scale field structure originating from the global properties of the flux emergence pattern. As described for the most active star in Section \ref{SubSec:ComparingTheInputMapsWithTheZDIMaps}, and as shown in Fig.~\ref{Fig:33_1992_ZDI}b, ZDI identifies a predominantly an azimuthal ring of negative polarity on the northern hemisphere. This field structure is highly axisymmetric and toroidal. The more active the star and the lower the inclination angle, the more the azimuthal toroidal ring dominates the Stokes~V profiles and ZDI struggles to recover the details of the relatively low contrast smaller-scale field on top of the azimuthal toroidal ring. This is likely a consequence of the maximum entropy regularisation within ZDI, which pushes to an image requiring the least amount of information in order to reconstruct the Stokes V profiles; the penalty for adding additional smaller-scale structure on top of the ring likely dominates over the much smaller improvement in the degree of fit to the input Stokes V profiles. This can result in the tendency of the azimuthal ring to dominate in the low inclination star with the highest activity level.

We want to highlight that the energy values and fractions for the $\ellsum\ = 7$ input maps are calculated for the whole surface. Due to the inclination angle ZDI will always miss parts of the southern hemisphere. Taking only the visible surface of the input maps into account would reduce the axisymmetric and toroidal fraction of the mean values per stellar model by at most $6\,\%$ and $2\,\%$ respectively, as well as the axisymmetric poloidal and the axisymmetric toroidal fraction by at most $5\,\%$ and $9\,\%$ respectively for the lower inclination angle $i = 20^{\circ}$. This correction would be even less for the higher inclination angles.

%
%
%
\begin{figure*}
\centering
\includegraphics[height=6.15cm, trim = {0 0 210 0}, clip]{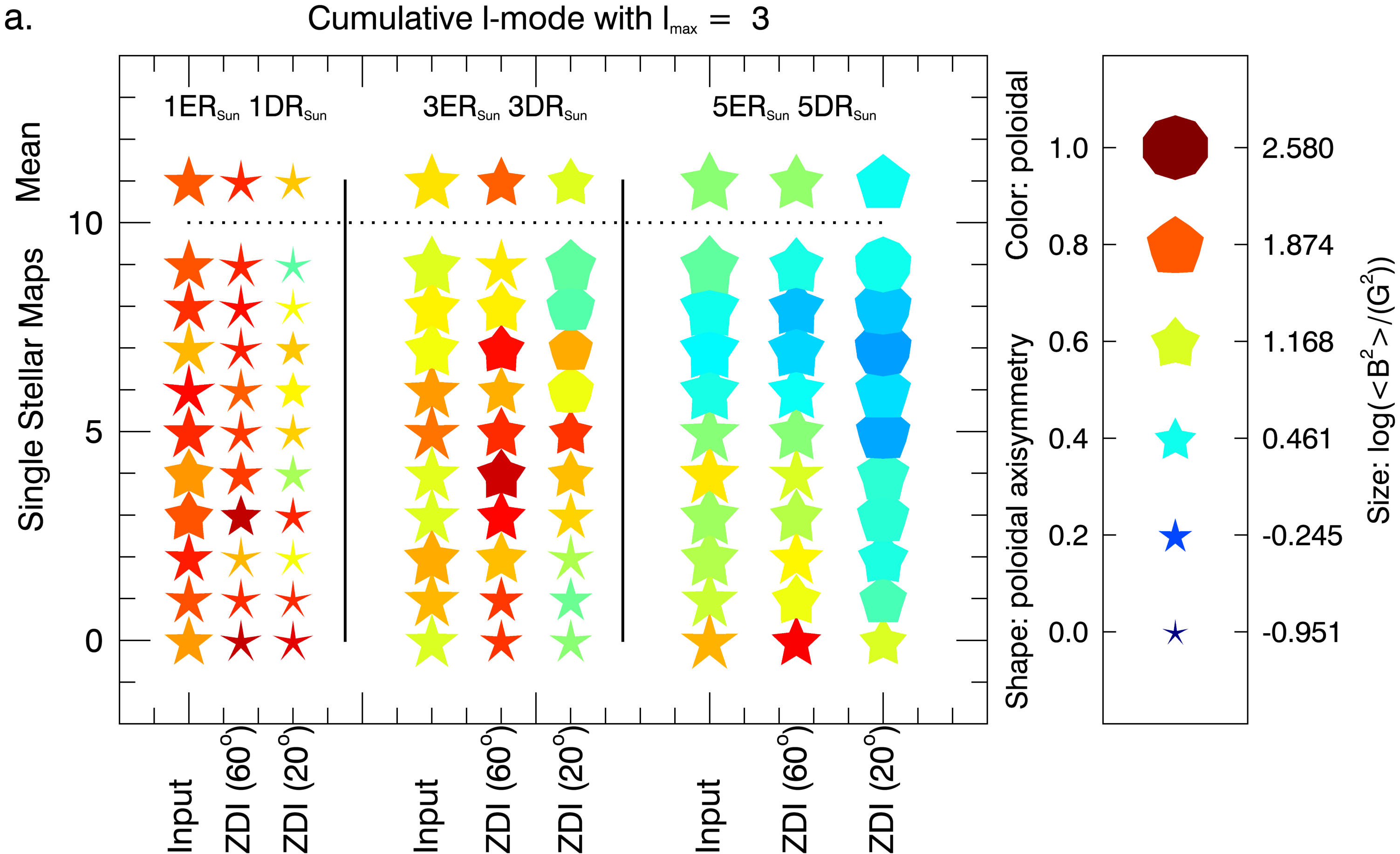}
\includegraphics[height=6.15cm]{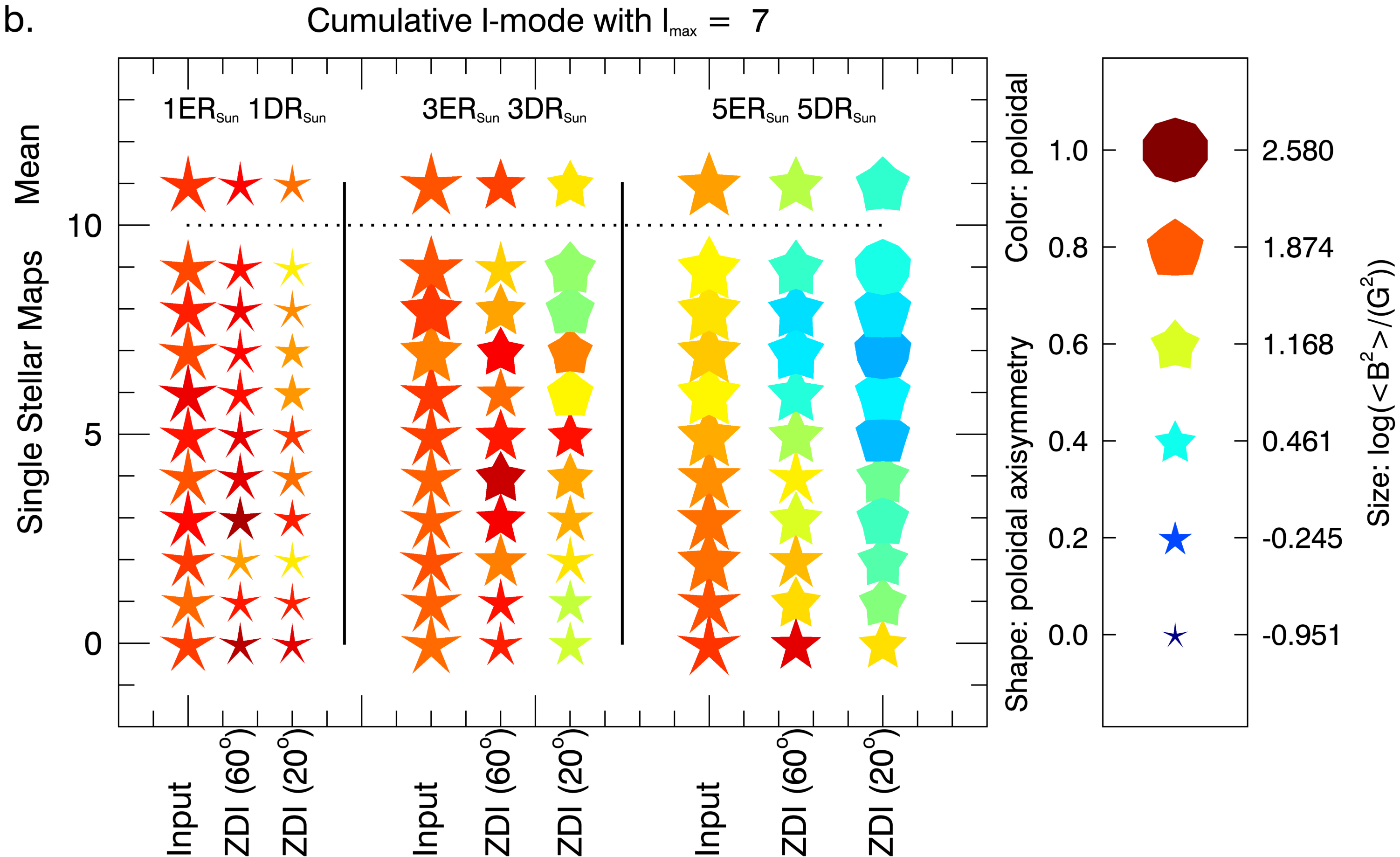} 
\caption{The properties of the large-scale field. The symbol size indicates the logarithmic magnetic energy $\log \langle B^2 \rangle$. The symbol colour indicates the poloidal fraction $f_{\mrm{pol}}$ and the shape the axisymmetry of the poloidal field $f_{\mrm{pol,axi}}$. We plot all ten maps per star below the dashed line and the mean above the dashed line. The different stellar models are separated by a solid black line. We display for each stellar model from left to right the input map, the ZDI reconstructed map for $i = 60^{\circ}$ and $i = 20^{\circ}$. The best agreement between the input and the ZDI reconstruction can be find at $\ellsum\ = 3$ (\textbf{a.}). Additionally, we show the maximal ZDI resolution with $\ellsum\ = 7$ (\textbf{b.}).
}
\label{Fig:Confusogram}
\end{figure*}

\citet[fig.~3]{Donati2009} summarised the large-scale field properties of the observed cool stars in a five dimensional diagram that is often use to compare different stellar magnetic maps (\citealt{Morin2008,Donati2011,Folsom2016,Hebrard2016}). 
Fig.~\ref{Fig:Confusogram} displays the input and ZDI reconstructed maps in this format. The left panel, Fig.~\ref{Fig:Confusogram}a, shows the large-scale field properties restricted to $\ellsum\ = 3$ and the right panel, Fig.~\ref{Fig:Confusogram}b, for $\ellsum\ = 7$;  we find that the large-scale field properties of the input and ZDI reconstructed maps are most similar for $\ellsum\ = 3$. The thick black lines separate the three different stellar models from each other. We plot the input and both ZDI reconstructed maps per stellar model from left to right. We show the ten single maps below and the average over them above the black dotted line. The symbol size relates to logarithmic total magnetic energy $\log \langle B^2 \rangle$, the symbol colour indicates the poloidal fraction $f_{\mrm{pol}}$ and the symbol shape displays the fraction of axisymmetric poloidal field $f_{\mrm{pol,axi}}$.

We see that the large-scale field morphology varies in the ten arbitrarily picked input maps for each stellar model, even on timescales that are much shorter than the long-term activity cycles. ZDI is sometimes able to recover these variations, see e.g. ZDI $i=60^{\circ}$ for the most active star ($\mrm{ER} = 5\,$\ERSun\ \ and $\mrm{DR} = 5\,$\DRSun\ ), however, this is especially hard for low inclinations. Comparing the large-scale field for $\ellsum\ = 7$, which is the highest mode ZDI is able to reconstruct, see Fig.~\ref{Fig:Confusogram}b, we see that the large-scale properties of the input and reconstructed maps fit less well. The reconstructed field strength is lower, and the more active the star, the more the poloidal fraction tends to be underestimated and the axisymmetry overestimated. A lower inclination angle enhances these trends. However, several individual maps for all stars and especially the averaged maps for the solar-like star and the more active star for $i=60^{\circ}$ are  in good agreement and show that ZDI is able to reconstruct the large-scale field morphology of solar-like stars for higher inclinations, especially on average.

\subsection{Recovering the energy distributions}

\begin{figure*}
\centering
\includegraphics[width = 0.49\textwidth ,clip]{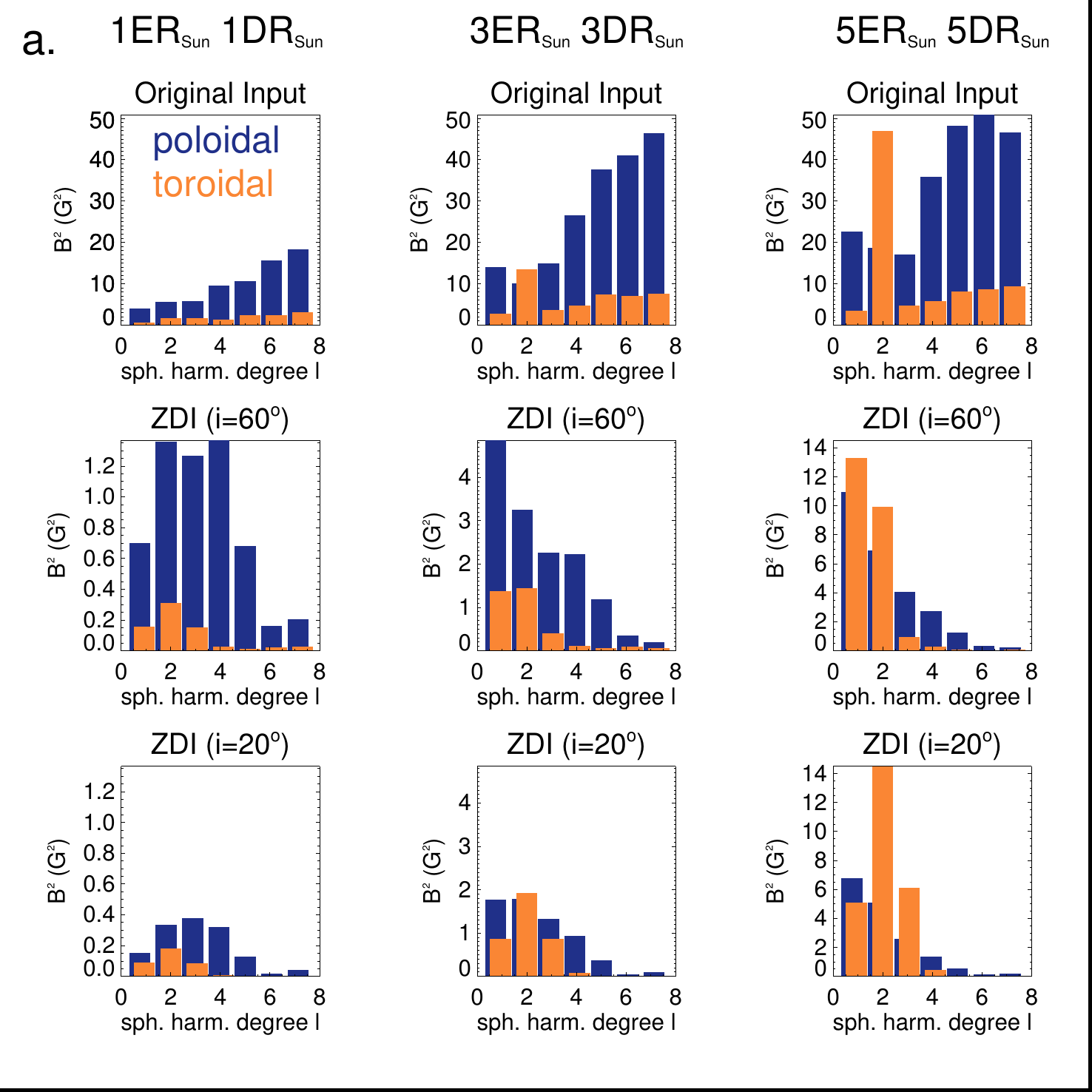} 
\includegraphics[width = 0.49\textwidth ,clip]{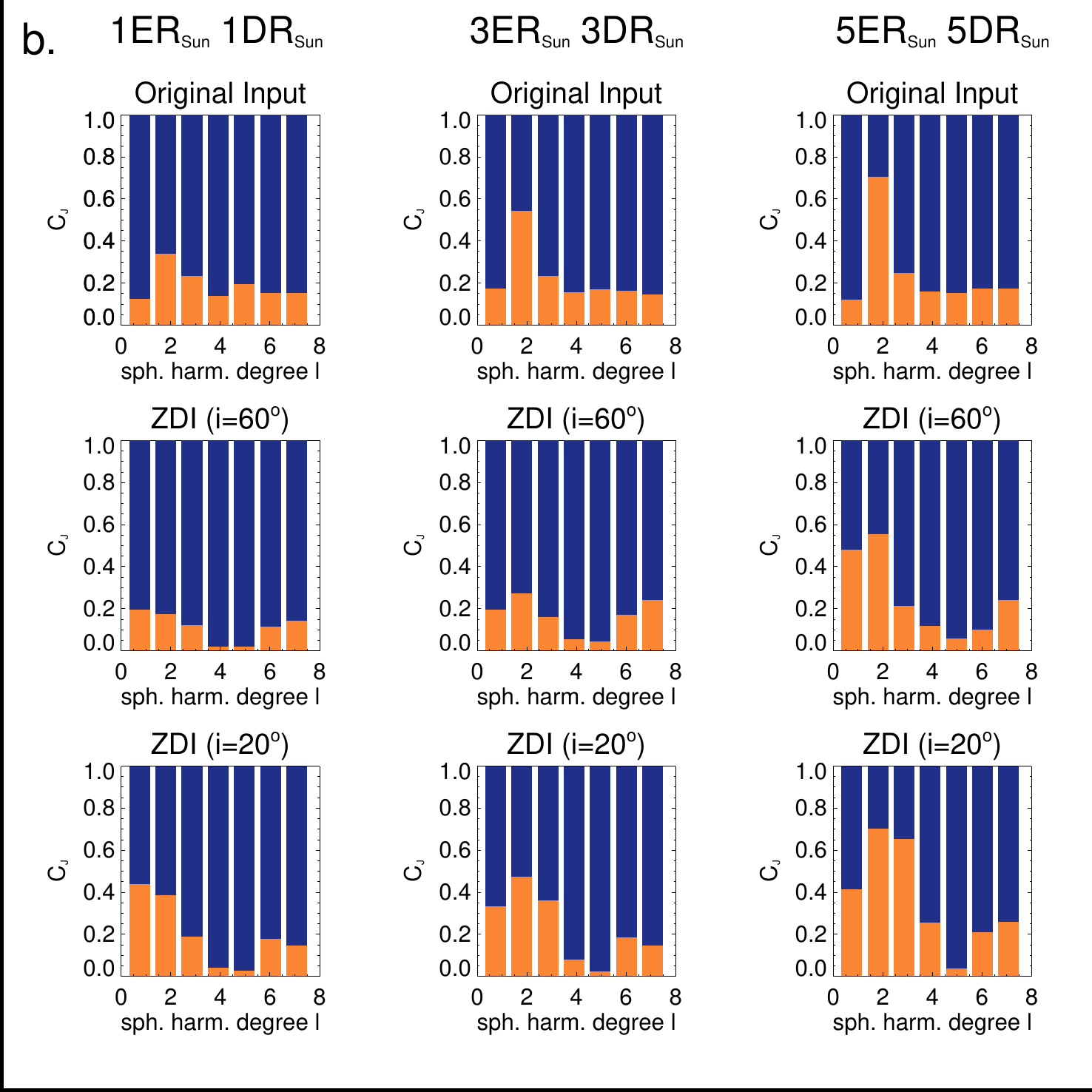} \\
\hspace{0.1mm}
\includegraphics[width = 0.49\textwidth ,clip]{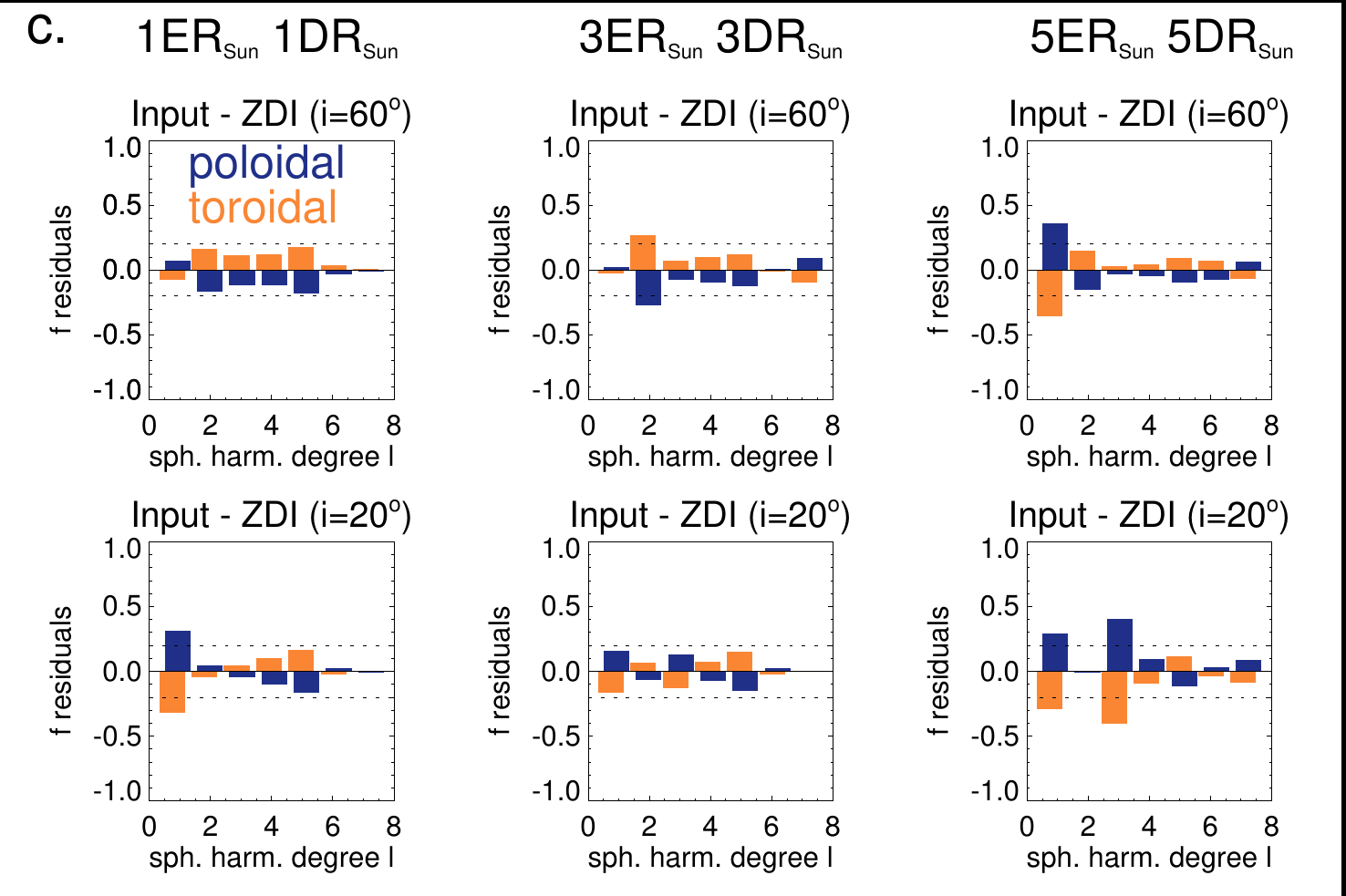} 
\includegraphics[width = 0.49\textwidth ,clip]{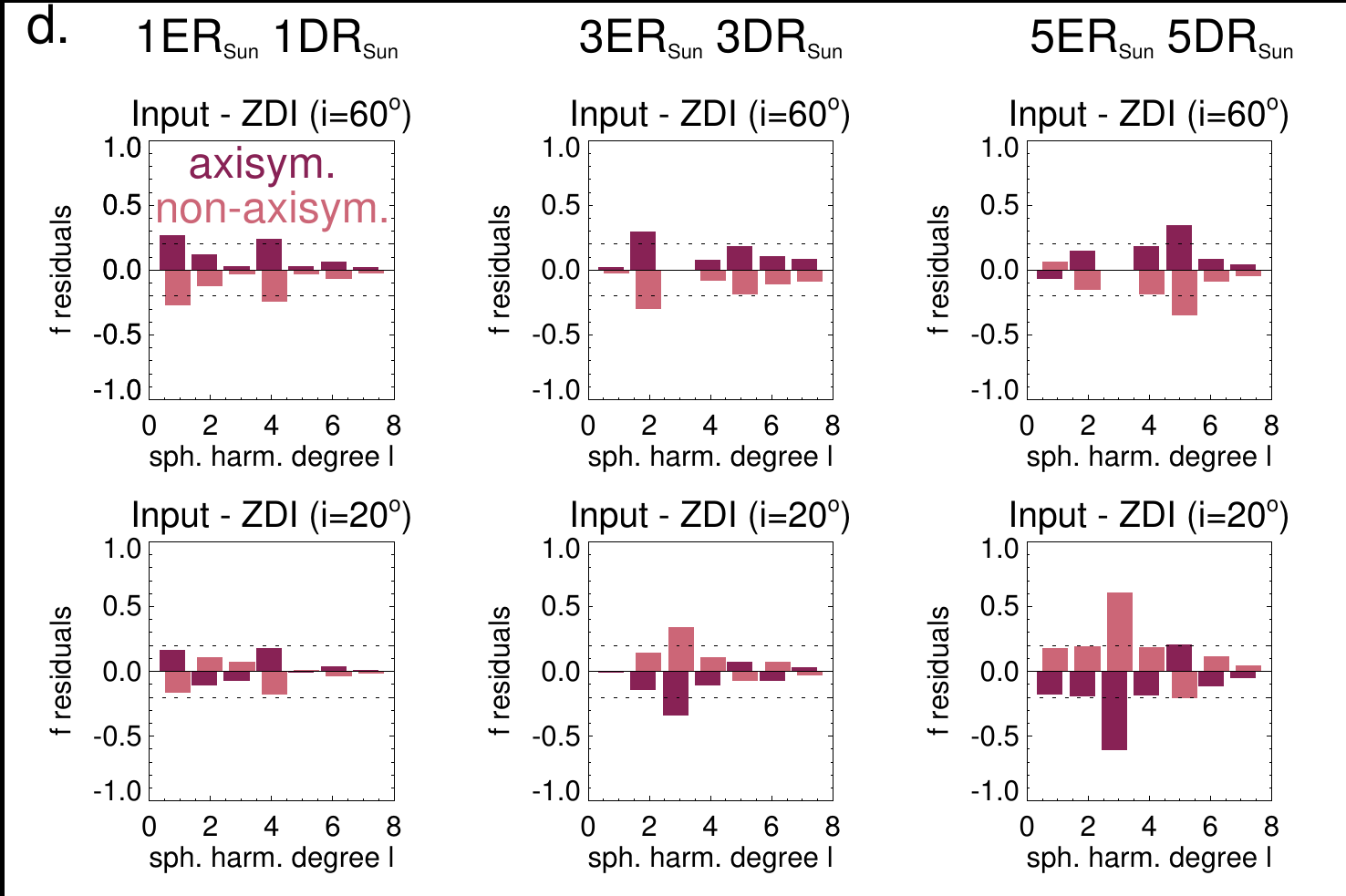}
\caption{The energy distribution (\textbf{a.}) and the fractions (\textbf{b.}) for the first seven $\ell$-modes of the toroidal (orange) and poloidal (plum) field component for the input maps (top row) and ZDI reconstructions for inclination $60^{\circ}$ and $20^{\circ}$ (2nd and 3rd row). On the bottom we plot the residuals (input - reconstruction) for both inclinations for the toroidal (orange) and poloidal (plum) fraction (\textbf{c.}) and for the axisymmetric (dark violet) and non-axisymmetric (rosa) field component (\textbf{d.}).
}
\label{Fig:BP_TorPolAxiNax}
\end{figure*}

We find that ZDI is not effective in reconstructing the magnetic energy distributions across individual  $\ell$-modes. Fig.~\ref{Fig:BP_TorPolAxiNax}a shows the energy distribution for the poloidal (plum bars) and toroidal (orange bars) field component. The columns present the three different stellar models (activity rises from left to right) and the rows show the first seven $\ell$-modes of the original input map and the ZDI reconstructed maps for both inclinations, respectively. We present the averaged results over the ten maps per stellar model and inclination.

For the  input maps, the magnetic energy increases with increasing $\ell$-mode. For the ZDI reconstructions, the magnetic energy mainly decreases with $\ell$-mode, i.e. the opposite trend. The reconstructions of the solar-like star show a peak in the poloidal energy at $\ell =2$ or 3, while for the more active stars the poloidal energy peaks at $\ell =1$. The input maps show that the toroidal energy becomes dominant at $\ell = 2$ the more active the star (see \citealt{Lehmann2018}). ZDI also reproduces the maximal toroidal energy at $\ell = 2$ most of the time, but not always. Also the magnetic energy values recovered by ZDI are much lower. The energy distributions of our ZDI reconstructions show a similar trend with $\ell$-modes found by observations of cool stars (see e.g. \citealt{Jeffers2014,Rosen2016}). We want to highlight again, that we are not pushing the solution towards or preferring any of the $\ell$-modes. For slowly rotating stars, we find that the maximum entropy implementation of ZDI naturally distributes energy over the allowed $\ell$-modes, with the energy tending to decrease with increasing $\ell$-mode. 

The picture improves if we compare the relative fractions of the field components, see Fig.~\ref{Fig:BP_TorPolAxiNax}b. The figure shows the cumulative total $C_J(\ell)$ of the poloidal (plum) and toroidal (orange) fraction, where
\begin{align}
C_J(\ell) &= \sum_j f_{j}(\ell) \\
f_{j}(\ell) &= \frac{\langle B^2_{\mathrm{j}}\rangle(\ell)}{\langle B^2_{\mathrm{tot}}\rangle(\ell)}, j \in \mrm{(pol, tor)} \\
\langle B^2_{\mathrm{tot}}\rangle(\ell) &= \langle B^2_{\mathrm{pol}}\rangle(\ell)+\langle B^2_{\mathrm{tor}}\rangle(\ell).
\end{align}
For the more active stars, ZDI is able to recover that the quadrupolar mode at $\ell =2$ has the highest fraction of toroidal field but it tends to overestimate the  toroidal fraction in the dipolar mode and underestimate the toroidal fraction for $\ell = 4,5$. The ZDI reconstructions show similar trends with $\ell$-modes for both inclinations, but the toroidal fraction is in general higher for lower inclination angles. 

For an easier comparison, we plot the residuals (subtracting the fraction of the ZDI reconstructed maps from the fraction of the simulated input maps) in Fig.~\ref{Fig:BP_TorPolAxiNax}c. The dashed lines indicate whether the mismatch is higher then $20\,\%$ for the fractions while using the same format as before. Positive values indicate that ZDI underestimates this field component and negative values that ZDI overestimates this field component. ZDI reproduces the input poloidal and toroidal fractions within a 20\,\% error for $\ell \geq 4$. For higher inclinations ZDI tends to underestimate the toroidal field especially for the less active stars, and overestimates the poloidal field ($f_{\mrm{tor}} = 1-f_{\mrm{pol}}$). For lower inclinations, ZDI tends to miss the fractions of the poloidal field of the dipolar and octopolar ($\ell = 1,3$) mode and  accordingly adds toroidal field.

We also analysed the other field components and see a similar picture: ZDI is not able to recover the magnetic energy distribution across ${\ell}$-modes, but recovers the fractions reasonably well. Fig.~\ref{Fig:BP_TorPolAxiNax}d, shows the axisymmetric field (dark violet) and the non-axisymmetric field (rosa). For the more active stars, we note that ZDI tends to underestimate axisymmetric field for the higher inclinations, and non-axisymmetric field for the lower inclinations. Looking at the residuals for the two components of the toroidal field: the azimuthal toroidal field and the meridional toroidal field component, we find that ZDI underestimates the meridional toroidal field by adding azimuthal toroidal field for lower inclination angles, while performing better at the higher inclination angles.

%
%
%

\section{Discussion and Conclusion}
\label{Sec:DiscussionConclusion}

First, we analysed the reconstruction abilities of four different magnetic field descriptions and the influence of $v_e \sin i$ and of the spatial resolution of the input map on the Stokes~V profiles.

We compared four different descriptions of the magnetic field, which are based on the spherical harmonics (see Section~\ref{sec:shdescription}). For the solar-like model it was possible to find single maps that could be reconstructed with a potential description ($\gamma_{\ell,m} = 0$, Eq.~\ref{Eq:B_tor}), but the majority of the solar-like model maps, as well as all maps of the more active stars, need a non-potential description ($\gamma_{\ell,m} \neq 0$). For a few maps, it was possible to set $\alpha_{\ell,m} = \beta_{\ell,m}$ (Eq.~\ref{Eq:B_rad}-\ref{Eq:B_mer}) which allows the extrapolation of non-potential fields to the outer atmospheres, but this approach reduces the degrees of freedom, and is only suitable for selected individual maps. The restriction of the description of the magnetic field by setting $\alpha_{\ell,m} = \beta_{\ell,m}$ and/or $\gamma_{\ell,m} = 0$ prevents the exact reconstruction of the magnetic field from a theoretical perspective. Only certain magnetic field morphologies are able to fit within these restrictions. To ensure that the reconstructed magnetic field is correct one should allow $\alpha_{\ell,m} \neq \beta_{\ell,m}$ and $\gamma_{\ell,m} \neq 0$ as this description is always able to construct the input magnetic field. This is especially important if one is interested in the azimuthal, meridional or toroidal field components. 
It is often impossible to discard single descriptions by fitting the Stokes~V profiles with the different magnetic field descriptions. \cite{Kochukhov2016a} showed that the Stokes~V profiles of the B0\,V star $\tau$\,Sco could be fitted equally well with three of the descriptors analysed here, but this resulted in three different vector magnetic field maps and magnetic field morphologies.

Further, we visualize the dependence of the Stokes~V profiles on $v_e \sin i$ and the spatial resolution of the input maps in Fig.~\ref{Fig:StokesIVFit}, Section~\ref{sec:vsinistokes}. For slow rotators like the Sun with $v_e \sin i \le 5\,\mrm{km/s}$, the Stokes~V profile does not change after including $\ellsum\ = 5-7$ to the input maps. This indicates that the observer is literally blind to magnetic field structures at smaller scales. These dependencies are widely known and discussed in the literature (see e.g. \citealt{Morin2008a}). However, there has not been a comprehensive exploration of the effects of this limited spatial resolution on a large set of input maps that include structures from  $\ellsum\ = 1-28$, in the $v_e \sin i$ range of $1.6-24\,\mrm{km/s}$.

We discuss in the following the ability of the ZDI technique to recover the large-scale field of 3D non-potential magnetic field simulations.

First of all, we want to highlight our result that the spherical harmonics decomposition provides an easy and fast method to estimate the large-scale magnetic field recoverable by ZDI of any vector magnetic field map. 
\cite{Yadav2015} presented that the ZDI reconstruction of a simulated vector magnetic field map of a more rapidly rotating fully convective star with $v_e \sin i = 20\,\mrm{km/s}$ can be mimicked by large-scale field spherical harmonics decomposition including $\ellsum\ = 10$. For solar-like stars, which are relatively slow rotators with $v_e \sin i \le 5\,\mrm{km/s}$, we show that the inclusion of $\ellsum\ = 5$ provides an adequate estimation of the large-scale magnetic field structure that would be observed via ZDI. The spherical harmonic decomposition\footnote{See \cite{Vidotto2016} for further details of the implementation.} enables one to determine the large-scale field of any vector magnetic field map, e.g. from solar-dynamo simulations, and to directly compare them with cool stars magnetic field observations, without running and applying a full ZDI routine. Caution is needed as these comparisons are better suited to the field morphologies and not the absolute magnetic field or energy values.

Secondly, we show that ZDI is able to reconstruct the main structures of the large-scale field of solar-like stars. The more active and highly inclined (i.e. less pole-on) the star, the better ZDI reconstructs the large-scale field of the three field components. Especially after averaging several maps, there is agreement in the large-scale field properties. 
The large-scale fields of solar-like stars can vary notably from one stellar rotation to another, even on timescales considerably shorter than the  activity cycle lengths. ZDI is generally able to follow these variations in the more active solar-like stars and high inclination angles, but there are some exceptions.  

We find that ZDI provides better results for higher inclination angles (i.e. more equator-on) than for lower inclination angles (pole-on). The effects of the different inclination angles are relatively small and in the range of the spread of the single maps per stellar model, e.g. see Fig.~\ref{Fig:EpolEtor}, but there are also definite trends, and some significant differences, e.g. see Fig.~\ref{Fig:AxiTorAxiPolAxiTor}. For lower inclination angles, less magnetic energy is recovered, especially less poloidal energy. For these pole-on views, less of the stellar surface is visible and as we always integrate over the whole surface, this could explain partly the lack magnetic energy for lower inclination angles, but it does not explain the lack of poloidal energy. 
Accordingly, the relative fraction of the toroidal field is higher and the axisymmetry of the total field and of the poloidal and toroidal component increases for lower inclinations. The axisymmetric toroidal component is overestimated by 40-50\,\%. 
Furthermore, we see trends with stellar activity and $v_e \sin i$ for the ZDI reconstructed maps. The more active the star (showing a few times higher flux emergence and differential rotation rates than the Sun) the higher the reconstructed toroidal energy is in connection with low inclinations angles and we see a higher fraction of axisymmetric and toroidal field in general. 

Both effects, inclination and activity, need to be discussed in the context of the solar flux emergence pattern. The global or large-scale structure of the small-scale solar flux emergence pattern is characterised by a strong quadrupolar $\ell =2$ toroidal azimuthal mode (= two rings of opposite polarity across the equator at mid latitudes), which is also highly axisymmetric. This large-scale component is needed to support the field that emerges from the active regions appearing at mid to low latitudes around the equator, which show the opposite polarity patterns in the different hemispheres. Furthermore, the active regions appear in a very axisymmetric pattern. The higher the flux emergence rate the more active regions are present at the same time and the stronger  the axisymmetric toroidal $\ell = 2$-mode (see Figs.\ref{Fig:11_1957_ZDI}-\ref{Fig:33_1992_ZDI}b). An increase in differential rotation decreases the poloidal field of the active regions, which further increases the toroidal fraction, see \cite{Lehmann2018}. The increase of the axisymmetry and toroidal fraction with stellar activity results therefore from the strengthening of the axisymmetric toroidal $\ell = 2$ mode, mainly due to the increase of active regions on the stellar surface following the solar flux emergence pattern. Observing these active stars at a lower inclination angle enhances this effect. With a near pole-on view, the southern hemisphere is not visible and most of the magnetic field originates from active regions which are now observed mainly on the limb while the polar region contains only very weak magnetic field. The structure that dominates the large-scale field is the northern magnetic ring of the axisymmetric toroidal $\ell = 2$ modes, which causes the overestimation of the toroidal and axisymmetric fraction by ZDI due to the line-of-sight effect on the solar flux emergence pattern.

This may also provide an explanation for the observed cool stars with Rossby numbers $\mrm{Ro} \le 1.0$ but stellar mass $M \ge 0.5\,\mrm{M_{\odot}}$, which show strong toroidal bands at low to mid latitudes. It could be that a strong $\ell=1$ or $\ell=2$ toroidal mode indicates well populated active latitude regions on these stars. However, these stars also have a much higher $v_e \sin i$, which is not covered by our simulations, and their significantly enhanced levels of flux emergence may lead to unknown flux distributions. We also note that strongly poloidal fields can still be found in ZDI reconstructions of slowly rotating solar-type stars (e.g. HD~146233, \citealt{Petit2008}, or HD~147513, \citealt{Hussain2016}).

We find that ZDI does not recover the magnetic energy distribution across individual $\ell$-modes well, neither in absolute values, nor in the relative trends. The energy distribution of the ZDI reconstructed maps (mainly decreasing energy with $\ell$-mode) is opposite to the energy distribution of the input maps (increasing energy with $\ell$-mode) and the energy values are much lower. Our ZDI reconstructions did not artificially add weights in favour of the low $\ell$-modes, i.e. all $\ell$-modes were defined to have equal weights. The trend to simpler structure in the ZDI reconstructions likely results from the maximum entropy implementation: as maximum entropy regularisation searches for the simplest field required to fit the set of Stokes profiles, it will tend to add structure into the lowest energy modes first. As it is an ill-posed problem, with an infinite number of energy distributions that can reproduce the same circularly polarised time-series, ZDI is unable to recover the  input distribution across $\ell$-modes. 

Nevertheless, ZDI is able to recover the fractions of the individual magnetic field components within $\sim 20\,\%$ for individual $\ell$-modes. For lower inclination angles and more active stars, we find an overestimation of the (azimuthal) toroidal and axisymmetric field due to the same reasons as discussed above. That ZDI is able to recover the field fractions to a satisfying level is very encouraging for the interpretation of the magnetic field morphologies of observed cool stars. We should be cautious in using the absolute values of the magnetic field but the fractions of the different field components are mostly reliable within $20\,\%$ for solar-like stars.

We find that ZDI is not able to reconstruct the correct levels of field strengths and energies. Even if we account for inclination instead of comparing with the entire surface input maps, the reconstructed field strengths and energies would only increase by a few percent. This is mainly due to the fitting of the Stokes profiles. The Stokes profiles can be fitted for the same $\chi^2$ with a range of magnetic field values for solar-like stars. As maximum entropy regularisation in ZDI searches for the lowest energy solution, the actual magnetic field can be underestimated and still yield an equally good fit to the Stokes V profiles (as defined by reduced $\chi^2$). We analyse the amount by which ZDI underestimates the field, by dividing the average magnetic field of the ZDI reconstructed map $\langle B_{\mrm{ZDI}}\rangle$ with the average magnetic field of the input map $\langle B_{\mrm{Inp}}\rangle$ per $\ell$-mode. There is a large variation across individual maps, leading to a range of values. The average magnetic field per $\ell$-mode is recovered in the range of $10-110\,\%$ depending on stellar activity, $\ell$-mode and inclination angle. In general, we see that with increasing activity level (and therefore increasing $v_e \sin i$) the recovered percentage increases. For example the dipolar mode $\ell = 1$ is recovered by $\langle B_{\mrm{ZDI}}\rangle/\langle B_\mrm{{Inp}}\rangle = 0.40 \pm 0.11$ for the solar-like star, by $\langle B_{\mrm{ZDI}}\rangle/\langle B_\mrm{{Inp}}\rangle = 0.58 \pm 0.12$ for the more active star and by $\langle B_{\mrm{ZDI}}\rangle/\langle B_\mrm{{Inp}}\rangle = 0.96 \pm 0.16$ for the most active star. The higher the $\ell$-mode the lower the $\langle B_{\mrm{ZDI}}\rangle/\langle B_\mrm{{Inp}}\rangle$ in general, but there are also a number of exceptions. The large spread and the high dependence on $v_e \sin i$ and inclination angle makes it impossible to provide a general correction factor for the underestimation of the magnetic field by ZDI especially if one keeps in mind that our analysed stars cover only a very small fraction of the observed cool stars. The non uniform dependency on the different $\ell$-modes can also be found by comparing the solar magnetograms of different solar observations as shown by \cite{Virtanen2017}.

%
%
%
\begin{figure}
\centering
\includegraphics[width = 0.49\textwidth ,clip]{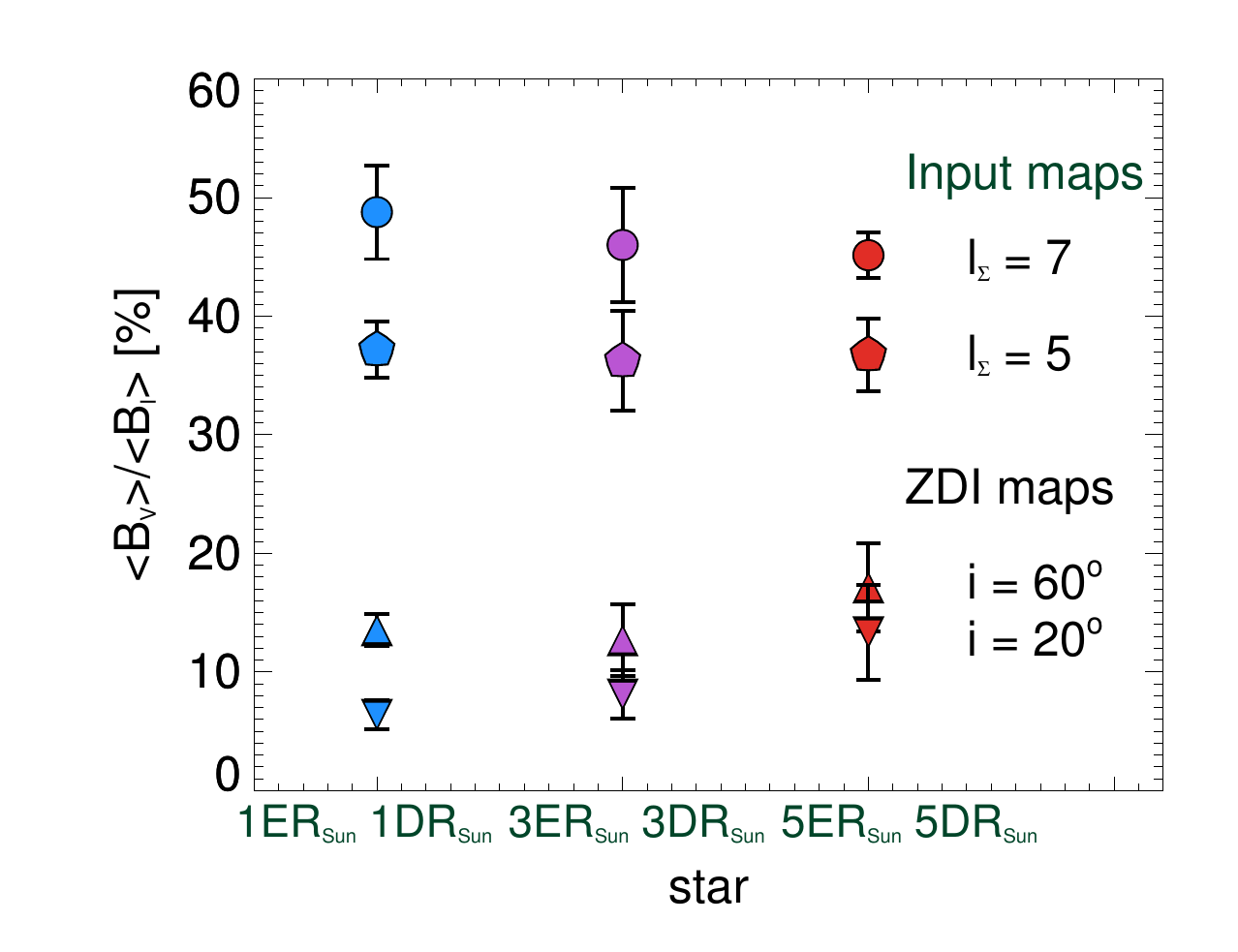} \\
\caption{The fraction of the mean flux density that is recovered by the large-scale field of the input maps for $\ellsum\ = 5$ and $7$ and the ZDI reconstructed maps. $\langle B_V\rangle/\langle B_I\rangle$ for the three different stars, where $\langle B_I\rangle$ is the unsigned mean magnetic flux density of the full resolved input map. For the circular and pentagon symbols $\langle B_V\rangle$ is equal to the large-scale field for $\ellsum\ = 7$ and $5$ of the input maps. For the normal and up-side-down triangular symbols $\langle B_V\rangle$ is equal to the ZDI reconstructed maps for the inclination angles $i=60^{\circ}$ and $i=20^{\circ}$.  
}
\label{Fig:BvBi}
\end{figure}

\cite{Reiners2009} and \cite{Morin2010} showed that the averaged magnetic field measured by ZDI $\langle B_V\rangle$ represents only $6-14\,\%$ of the total averaged magnetic field measured by Zeeman broadening $\langle B_I\rangle$ for M~dwarfs. Recently, \cite{Vidotto2018} showed that the large-scale field of the synoptic maps of solar activity cycle 24 includes only $\approx 10-20\,\%$ of the total field most of the time. In Fig.~\ref{Fig:BvBi}, we present our analysis of $\langle B_V\rangle/\langle B_I\rangle$ for the three stellar models using the same colour schema as in Fig.~\ref{Fig:EpolEtor}. The circles and pentagons represent the averaged large-scale field $\langle B_V\rangle$ of the input maps including $\ellsum\ = 7$ and $5$. For the normal and upside-down triangle is the averaged large-scale field $\langle B_V\rangle$ of the ZDI reconstructed maps for inclination $i=60^{\circ}$ and $i=20^{\circ}$. We calculate $\langle B_I\rangle$ from the fully resolved input map, so that it corresponds to the total averaged magnetic field. The symbols in Fig.~\ref{Fig:BvBi} represent the mean over the ten maps per stellar model and the error bars indicate the corresponding standard derivation.

We find that the large-scale field of the input maps with $\ellsum\ = 7$ and $\ellsum\ = 5$ comprises $\approx$\,$47\,\%$ and $\approx$\,$37\,\%$ of the total average magnetic field, independent of the activity level of the star. For the ZDI reconstruction, we find that ZDI recovers $\approx$\,$12-17\,\%$ for higher inclination angles and $\approx$\,$6-13\,\%$ for lower inclination angles. Next to this trend in inclination we see a trend with stellar activity. The more active the star, the higher its $\langle B_V\rangle/\langle B_I\rangle$. Both effects result from the dependence of the ZDI resolution on $v_e \sin i$. The more active the star, and the higher its inclination angle, the higher its $v_e \sin i$, and therefore the resolution and the field reconstructed by ZDI. If we account for the obscuration effects of stellar inclination, we see that the $\langle B_V\rangle/\langle B_I\rangle$ mean values per stellar model for the input maps would be reduced by less than $\approx$\,$6\,\%$ and the $\langle B_V\rangle/\langle B_I\rangle$ for the ZDI maps increases by less than $\approx$\,$5\,\%$ for the lower inclination angle of $i = 20^{\circ}$ and even less for $i = 60^{\circ}$.
It is still remarkable that our $\langle B_V\rangle/\langle B_I\rangle$ results for solar-like stars are similar to those based on observations of more active M~dwarfs \citep{Reiners2009,Morin2010}.

\section{Summary}
\label{Sec:Summary}

We used 3D non-potential magnetic field simulation as points of reference to analyse the reliability of the ZDI technique when applied to slowly rotating, solar-like stars. Furthermore, we examined four different descriptions of the magnetic field based on spherical harmonics, and the dependency of the Stokes V profiles on both the spatial resolution and $v_e \sin i$. We found that ZDI is able to recover the main structures of the large-scale field morphology but with approximately one order of magnitude less magnetic energy. We compare the ZDI reconstructed maps with the whole surface input maps to provide realistic estimates what ZDI is able to recover. However, comparing the ZDI reconstructed maps with input maps only taken the visible surface into account would change the recovered property values only by a few percent.
The following list summarises our main conclusions:

\begin{itemize}
\item Only the non-potential magnetic field description with $\alpha_{\ell,m} \neq \beta_{\ell,m}$ and $\gamma_{\ell,m} \neq 0$ guarantees the correct reconstruction of the field, especially if one is interested in the azimuthal, meridional or toroidal field component. However, certain magnetic field morphology configurations can be recovered with the other spherical harmonic prescriptions, which place further restrictions on the spherical harmonic coefficients.
\item The Stokes~V profiles are insensitive to magnetic structures smaller than $\ellsum\ =5-7$, which corresponds to an angular resolution of $\theta \approx 36-25^{\circ}$, for slow rotators like the Sun with $v_e \sin i \le 5\,\mrm{km/s}$.
\item The spherical harmonic decomposition provides a fast and easy way to estimate the large-scale field of highly resolved vector magnetic field maps, which is recoverable using ZDI. We recommend using $\ell$-modes up to $\ellsum\ = 5$ for slow rotators.
\item ZDI is able to recover the main magnetic structures of the large-scale field for solar-like stars. The large-scale field properties are reasonably well recovered, especially after averaging over several maps. 
\item The large-scale field of solar-like stars can change significantly from one stellar rotation to another (even without accounting for activity cycles), and ZDI can follow these variations in the more active stars with higher inclination angles.
\item ZDI is affected by the inclination of the star, with higher inclination angles (more equator-on views) providing better results. The lower the inclination angle, the higher the fraction of toroidal and axisymmetric field for solar-like stars with solar flux emergence patterns. 
\item ZDI is affected by the rotation period and activity of the star. Within our sample the faster the star rotates, the higher is the number of resolution elements extracted from the Stokes profiles and therefore more magnetic field is recovered. The more active the star, the higher the toroidal and axisymmetric fraction for solar-like stars with solar flux emergences patterns.
\item The energy distributions across individual $\ell$-modes are not recovered by ZDI with a maximum entropy ansatz. However, the fractions of the different field components per $\ell$-mode are mostly recovered within an error of $20\,\%$. 
\item In general ZDI recovers less magnetic energy and magnetic field strength than expected from the large-scale field of the input maps. The average magnetic field per $\ell$-mode  detected by ZDI can range between $10-110\,\%$ of the corresponding average magnetic field of the same $\ell$-mode of the input map. Due to the large variations depending on $\ell$-mode, inclination, stellar activity and rotation, it is not possible to apply a fixed correction factor for the ZDI-reconstructed magnetic fields.
\item The averaged magnetic field detected by ZDI is $6-17\,\%$ of the average magnetic field of the fully resolved maps for solar-like stars. Shorter rotation periods and higher inclination angles increase the fraction.
\end{itemize}

We want to highlight that our focus for testing ZDI lies on ZDI's ability to recover the large-scale field structures of solar-like stars and their field properties. Our tests are therefore performed under optimal conditions, assuming a high SNR, and evenly spaced, well-sampled spectral time series. We also assume that the magnetic field maps are stable over the course of one stellar rotation. We will present the results of further tests including varying levels of SNR, phase coverage, and evolving magnetic field maps in an upcoming study.

\section*{Acknowledgements}

LTL acknowledges support from the Scottish Universities Physics Alliance (SUPA) prize studentship and the University of St Andrews Higgs studentship as well as for the support of the SUPA and the European Southern Observatory (ESO) for a 8-weeks short term visit at the ESO Headquarters in 2017 and the ESO for supporting a second visit (1-month) in 2018. Further, LTL wants to thank the Royal Astronomical Society (RAS) for a travel grant to visit the Cool Stars 20 Conference in Boston to present this work. We also warmly thank the IDEX initiative at Universite Ferale Toulouse Midi-Pyr\'enees (UFTMiP) for funding the STEPS collaboration programme between IRAP/OMP and ESO and for allocating a Chaire dAttractivite to GAJH, allowing her to regularly visit Toulouse and work on the application of ZDI techniques. DHM would like to thank the UK STFC for financial support.
The authors would like to thank Jean-Francois Donati and Pascal Petit for useful suggestions that have helped improve this study.




\bibliographystyle{mnras}
\bibliography{Lehmann2018ObservingTheSimulations3} 



\appendix

\section{additional figures}

\begin{figure*}
\raggedright
\large
\hspace{1.2cm}\textbf{Input map}\hspace{1.9cm}\textbf{Potential $\alpha = \beta$}\hspace{1.4cm}\textbf{Potential $\alpha \neq \beta$}\hspace{1.7cm}\textbf{Original map}\\
\hspace{0.9cm}(Visible surface) \hspace{9.5cm} (Complete surface)\\
\centering
\includegraphics[angle=0,width = \textwidth ,clip]{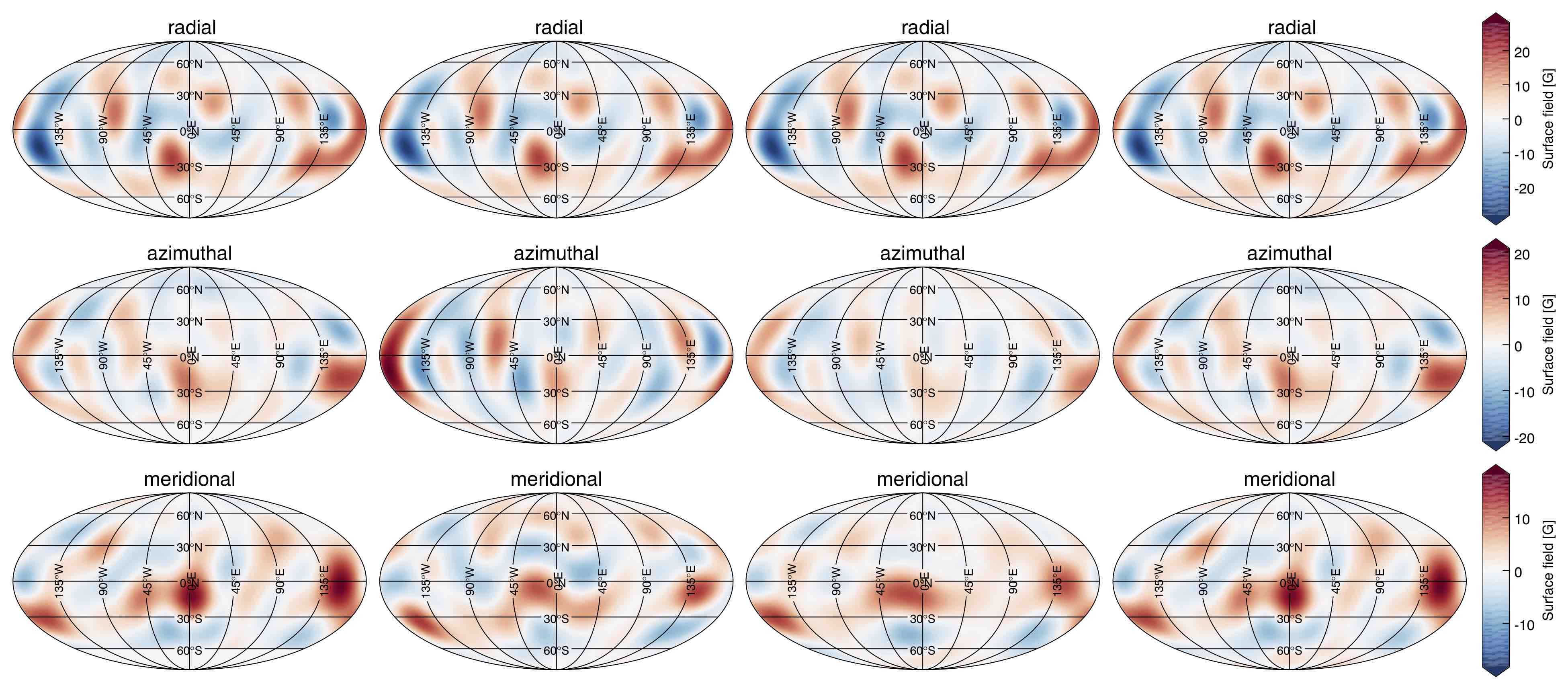}
\caption{The Mollweide projected magnetic field maps of the different descriptions of the magnetic field for the solar-like star ($\mrm{ER} = 1\,$\ERSun\ \ and $\mrm{DR} = 1\,$\DRSun\  ). Form \textit{left to right}: the input map including $\ell_{\Sigma} = 7$  and truncated latitudes corresponding to an inclination of $i=60^{\circ}$, the potential ($\alpha_{\ell m} = \beta_{\ell m}$) reconstruction, the potential ($\alpha_{\ell m} \neq \beta_{\ell m}$) reconstruction and the original simulated maps for $\ell_{\Sigma} = 7$. The same format as the Fig.~\ref{Fig:DecompMap11_1930} is used.}
\label{Fig:DecompMap11_2065}
\end{figure*}

\begin{figure*}
\raggedright
\large
\hspace{0.7cm}\textbf{Input map}\hspace{1.15cm}\textbf{Potential $\alpha = \beta$}\hspace{0.7cm}\textbf{Potential $\alpha \neq \beta$}\hspace{0.3cm}\textbf{Non-potential $\alpha = \beta$}\hspace{0.5cm}\textbf{Original map}\\
\hspace{0.4cm}(Visible surface) \hspace{10.6cm} (Complete surface)\\
\centering
\includegraphics[angle=0,width = \textwidth ,clip]{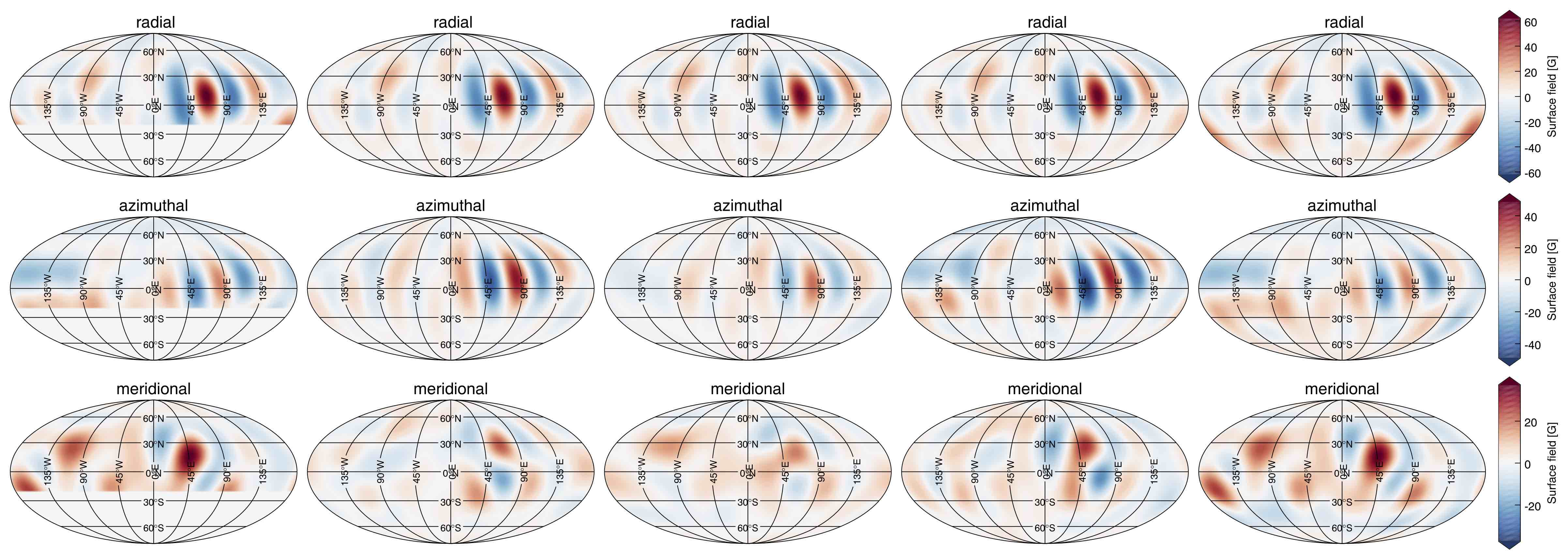}
\caption{The Mollweide projected magnetic field maps of the different descriptions of the magnetic field for the more active star ($\mrm{ER} = 3\,$\ERSun\ \ and $\mrm{DR} = 3\,$\DRSun\  ). Form \textit{left to right}: the input map including $\ell_{\Sigma} = 7$  and truncated latitudes corresponding to an inclination of $i=20^{\circ}$, the potential ($\alpha_{\ell m} = \beta_{\ell m}$) reconstruction, the potential ($\alpha_{\ell m} \neq \beta_{\ell m}$) reconstruction, the non-potential ($\alpha_{\ell m} = \beta_{\ell m}$) reconstruction and the original simulated maps for $\ell_{\Sigma} = 7$. The same format as the Fig.~\ref{Fig:DecompMap11_1930} is used.}
\label{Fig:DecompMap33_1992}
\end{figure*}

Fig.~\ref{Fig:DecompMap11_2065} and \ref{Fig:DecompMap33_1992} show additional examples for the potential $\alpha \neq \beta$ and the non-potential $\alpha = \beta$ field descriptions for Section~\ref{sec:shdescription}.

Fig.~\ref{Fig:VsiniCorrCoeff_20} shows additional to Fig.~\ref{Fig:VsiniCorrCoeff} in Section~\ref{sec:vsinistokes} the correlation coefficients between two successive $\ellsum\ $-modes for the three different stellar models (left to right) for the lower inclination $i=20^{\circ}$.

%
%
%

%
%
%

%
%

%
%
%
\begin{figure*}
\centering
\includegraphics[angle=0,width = \textwidth, trim={0 5cm 0 0} ,clip]{StokesVCorrCoeff.pdf}
\caption{The correlation coefficient $C_{\ellsum\ }$ between two successive $\ellsum\ $-modes for the three different stars (left to right) and an inclination angle of $i=20^{\circ}$. The colour of the curves indicates the applied rotation period, see Fig.~\ref{Fig:VsiniCorrCoeff} \textit{right}. }
\label{Fig:VsiniCorrCoeff_20}
\end{figure*}


\bsp	
\label{lastpage}
\end{document}